\documentclass[journal]{IEEEtran}
\ifCLASSINFOpdf
\else
\fi

\usepackage[utf8]{inputenc}

\usepackage{array}
\usepackage{subfigure}
\usepackage{epsfig,pifont}
\usepackage{epstopdf}
\usepackage{framed,multirow}
\usepackage{graphicx}
\usepackage{amssymb}
\usepackage{latexsym}

\usepackage{textcomp,booktabs}
\usepackage[usenames,dvipsnames]{color}
\usepackage{colortbl}
\usepackage{bm}
\usepackage{amsfonts,amssymb}
\usepackage{amsmath}
\usepackage{threeparttable}

\usepackage{color}

\definecolor{mygray}{gray}{.9}
\definecolor{mypink}{rgb}{.99,.91,.95}
\definecolor{mycyan}{cmyk}{.3,0,0,0}

\begin{document}
%
\title{Explainable Machine Learning based Transform Coding for High Efficiency Intra Prediction}
%
%
%

\author{Na~Li,
        Yun~Zhang,~\IEEEmembership{Senior Member,~IEEE,}
        C.-C. Jay~Kuo,~\IEEEmembership{Fellow,~IEEE}
\thanks{Na Li and Yun Zhang are with the Shenzhen Institutes of Advanced Technology, Chinese Academy of Sciences, Shenzhen, China, e-mail:\{na.li1, yun.zhang\}@siat.ac.cn}
\thanks{C.-C. Jay~Kuo is with Ming Hsieh Department of Electrical Engineering, University of Southern California, Los Angeles, California, USA, email:cckuo@sipi.usc.edu.}
 }

%
%

\markboth{}
{Shell \MakeLowercase{\textit{et al.}}: Bare Demo of IEEEtran.cls for IEEE Journals}
%



\maketitle

\begin{abstract}
 Machine learning techniques provide a chance to explore the coding performance potential of transform. In this work, we propose an explainable transform based intra video coding to improve the coding efficiency. Firstly, we model machine learning based transform design as an optimization problem of maximizing the energy compaction or decorrelation capability. The explainable machine learning based transform, i.e., Subspace Approximation with Adjusted Bias (Saab) transform, is analyzed and compared with the mainstream Discrete Cosine Transform (DCT) on their energy compaction and decorrelation capabilities. Secondly, we propose a Saab transform based intra video coding framework with off-line Saab transform learning. Meanwhile, intra mode dependent Saab transform is developed. Then, Rate-Distortion (RD) gain of Saab transform based intra video coding is theoretically and experimentally analyzed in detail. Finally, three strategies on integrating the Saab transform and DCT in intra video coding are developed to improve the coding efficiency. Experimental results demonstrate that the proposed 8$\times$8 Saab transform based intra video coding can achieve Bj$\o$nteggard Delta Bit Rate (BDBR) from -1.19$\%$ to -10.00$\%$ and -3.07$\%$ on average as compared with the mainstream 8$\times$8 DCT based coding scheme.
\end{abstract}


\IEEEpeerreviewmaketitle

\section{Introduction}
\label{Intr}

\IEEEPARstart{V}{ideo} data contributes the most in the increasing of data volume in the era of big data due to its realistic representation and wide application. Commercial broadcasting for video resolutions are expected to be extended from 4K Ultra-High Definition (UHD) to 8K UHD in the near future. Meanwhile, High Dynamic Range (HDR), holograph Three-Dimension (3D) and Virtual Reality (VR) videos boost by the attractiveness in providing realistic, 3D and immersive visual experiences. In addition, the usage of these video applications is striving with increasing number of video devices connected to the internet or Internet of Things (IoT), e.g., TV, laptops, smartphones, surveillance cameras, drones, etc.. Along the increase of both the usage and quality of videos, the volume of global video data doubles every two years, which is the bottleneck for video storage and transmission over network. In the development of video coding standards, from MPEG-2, H.264/Advanced Video Coding (AVC), H.265/High Efficiency Video Coding (HEVC) to the latest Versatile Video Coding (VVC)\cite{Bross2019}, the video compression ratio is doubled almost every ten years. Although researchers keep on developing video coding techniques in the past decades, there is still a big gap between the improvement on compression ratio and the volume increase on global video data. Higher coding efficiency techniques are highly desired.

In the latest three generations of video coding standards, hybrid video coding framework has been adopted, which is composed of predictive coding, transform coding and entropy coding. Firstly, predictive coding is to remove the spatial and temporal redundancies of video content on the basis of exploiting correlation among spatial neighboring blocks and temporal successive frames. Higher prediction accuracy leads to less residuals to be encoded, and thus leads to higher compression ratio. The predictive coding can be classified as intra prediction and inter prediction based on its reference pixels from spatial or temporal domain. Secondly, transform coding that mainly consists of transform and quantization is to transform residuals from predictive coding to a spectral domain, and then quantize the spectral coefficients to further exploit spatial and perceptual redundancies. For example, Human Vision System (HVS) is generally more sensitive to low frequency than high frequency signals, where larger quantization scales could be given. Finally, entropy coding exploits the statistical property of transformed coefficients so as to approach its entropy. Generally, it encodes symbols of higher probability with less bits whereas encodes symbols of lower probability with more bits. In this paper, we focus on developing an explainable machine learning based transform coding that improves the coding efficiency in the hybrid video coding framework.

Karhunen-Lo$\acute{e}$ve Transform (KLT) is an ideal transform for energy compaction and decorrelation, which requires to calculate an autocorrelation matrix for each source input block. In video coding, the autocorrelation matrix shall be encoded and transmitted associated with transformed coefficients, which brings an additional bit rate overhead while using KLT in video coding. There are cases the autocorrelation matrix are derived and stored without transmission. The outstanding energy compaction and decorrelation capabilities of KLT attract researchers to study data-driven transform.
Dvir {\em et al.} \cite{Dvir2020} constructed a new transform from an eigen-decomposition of a discrete directional Laplacian system matrix. Lan {\em et al.} \cite{Lan2018} trained one dimensional KLT through searching patches similar with the current block from reconstructed frames with computational overhead. As a derivation of the secondary transforms, Koo {\em et al.} \cite{Koo2019} learned non-separable transforms from both video sequences and still images out of the Call for Proposals (CfP), and adopted five of the transforms with the best Rate-Distortion (RD) cost in the reference encoder as the final non-separable transform.
Cai {\em et al.} \cite{Cai2015} proposed to only estimate the residual covariance as a function of the coded boundary gradient, considering prediction is very sensitive to the accuracy of the prediction direction in the image region with sharp discontinuities.
Wang {\em et al.} \cite{Wang2019} proposed to optimize transform and quantization together with RD optimization (RDO). Zhang {\em et al.}\cite{Zhang2020} designed a high efficient KLT based image compression algorithm.
Graph Based Transform (GBT) was proposed as a derivation of the traditional KLT in \cite{Egilmez2020}, which incorporates Laplacian with structural constraints to reflect the inherent model assumptions about the video signal. Arrufat {\em et al.} \cite{Arrufat2014} designed a KLT based transform for each intra prediction mode in Mode-Dependent Directional Transform (MDDT). Takamura {\em et al.}\cite{Takamura2013} proposed the non-separable mode-dependent data-dependent transform and create offline 2D-KLT kernels for each intra prediction modes and for each Transform Unit (TU) sizes.
In these recent studies, researches focused on generating autocorrelation matrix for the data-dependent KLT and optimizing the data-dependent KLT with the constrained autocorrelation matrix. It is not only computational difficult to estimate or encode the autocorrelation matrix for dynamic block residuals, but also memory consuming to store the transform kernels offline.

Discrete Cosine Transform (DCT) performs similarly to KLT on energy compaction when the input signal approximates Guassian distribution. Due to its good energy compaction capability and relative low complexity, DCT has been widely used in video coding standards, including MPEG-1, MPEG-2, MPEG-4, H.261, H.262 and H.263. H.264/AVC and later coding standards adopted Integer DCT (ICT) to approximate float-point DCT calculation for lower complexity and hardware cost. ICT was computed exactly in integer arithmetic instead of the float-point computation and multiplications were replaced with additions and shifts. Since DCT transform kernels are fixed and difficult to adapt to all video contents and modes, advanced DCT optimizations are proposed to improve the transform coding efficiency through jointly utilizing multiple transforms and RDO \cite{Zhao2012}. For intra video coding in HEVC, an integer Discrete Sine Transform (DST) \cite{ITU2013} was further applied to 4$\times$4 luminance (Y) block residuals.
Han {\em et al.} \cite{Han2012} proposed a variant of the DST named Asymmetric DST (ADST) \cite{Han2012} regarding to the prediction direction and boundary information.

Furthermore, due to the diversity of video contents and their distributions, multiple transforms from DCT/DST families were jointly utilized other than one transform to enhance the coding efficiency. Zhao {\em et al.} \cite{Zhao2019} presented the Enhanced Multiple Transform (EMT) by selecting the optimal transform from multiple candidates based on the source properties and distributions. EMT is intra mode dependent where DCT/DST transform kernels are selected based on the intra direction. As the coding efficiency of EMT comes with the cost of higher computational complexity at the encoder side, Kammoun {\em et al.} \cite{Kammoun2018} proposed an efficient pipelined hardware implementation. In 2018, EMT was simplified as Multiple Transform Selection (MTS) and adopted in the VVC by Joint Video Expert Team (JVET) \cite{Zhao2018CE6}, which consists of experts from ISO/IEC MPEG and ITU-T VCEG. Zeng \emph{et al.} \cite{Zeng2008} presented a two-stage transform framework, where coefficients at the first stage produced by all directional transforms were arranged appropriately for the secondary transform. Considering that multi-core transforms and non-separable transforms can capture diverse directional texture patterns more efficiently, EMT \cite{Zhao2019} and Non-Separable Secondary Transform (NSST) \cite{Zhao2018-2} were combined to provide better coding performance in the reference software of the next generation video coding standard.
Zhang \emph{et al.} \cite{ZhangY2020} presented a method on Implicit-Selected Transform (IST) to improve the performance of transform coding for AVS-3. Pfaff \emph{et al.} \cite{Pfaff2020} applied mode dependent transform with primary and secondary transforms to improve transform coding in HEVC.
Park \emph{et al.} \cite{Park2019} introduced fast computation methods of N-point DCT-V and DCT-VIII. Garrido \emph{et al.} \cite{Garrido2019} proposed an architecture to implement transform types for different sizes.
DCT is a pre-defined and fixed transform to approach KLT's performance for Guassian distributed source. However, the Guassian distributed source assumption cannot be always guaranteed due to the diversity of video contents and variable block patterns, which enlarge the gap between coding efficiency of using DCT and the best KLT. Video content with complex textures are difficult to follow Gaussian distribution. In addition, using multiple transform kernels for different source assumptions and selecting the optimal one with RDO significantly increase the coding complexity. 

Machine learning based transform is a possible solution to have a good trade-off between data dependent KLT and fixed DCT and improve the video coding performance. Lu {\em et al.}\cite{Lu2020} utilized non-linear residual encoder decoder network implemented as Convolutional Neural Network (CNN) to replace the linear transform. Parameters of these CNN are determined by backpropagation which is not algorithmic transparency, short of human comprehensibility, not robust to perturbations and hard to be improved layer by layer. To handle these problem, explainable machine learning is a possible solution to improve the interpretation, scalability and robustness of learning. Therefore, Kuo {\em et al.} \cite{Kuo2019} proposed an interpretable feedforward design, noted as  Subspace Approximation with Adjusted Bias (Saab) transform, which is statistics-centric and in unsupervised manner. Motivated by the analyses on nonlinear activation of CNN in \cite{Kuo2016,Kuo2017}, Kuo {\em et al.} \cite{Kuo2018} proposed the Subspace Approximation with Augmented Kernels (Saak), where each transform kernel was the KLT kernel augmented with its negative so as to resolve the sign confusion problem. The sign confusion problem was solved by shifting the transform input to the next layer with a relatively large bias in Saab transform \cite{Kuo2019}.
As the explainable machine learning method, Saab transform\cite{Kuo2019}, interpreted the cascaded convolutional layers as a sequence of approximating spatial-spectral subspaces.
The data-dependent, multi-stage and non-separable Saab transform was proposed to interpret CNN for recognition tasks, which also has good energy compaction capability. In \cite{Li2019}, Saab transforms were learned from video coding dataset and have potentials to outperform DCT on energy compaction capability for variable block size residuals from intra prediction. However, the coding performance of Saab transform for intra coding need to be further investigated. Saab transform based intra video coding for specific block size could be reproduced for other block sizes under the same methodology. Therefore, we mainly explore 8$\times$8 Saab transform for intra predicted 8$\times$8 block residuals to illustrate the coding performance potential of Saab transform.

In this work, we propose the Saab transform based intra video coding to improve the coding efficiency. The paper is organized as follows. Saab transform and its transform performances are analyzed in Section \ref{sec:dataDrivenTrCodec}. A framework of Saab transform based intra video coding and intra mode dependent Saab transform are illustrated in Section \ref{sec:frmwork}. Then, the RD performances and computational complexity are analyzed in Section \ref{sec:RDModelForTr}. Extensive experimental results and analyses are presented in Section \ref{Sec:Exp}. Finally, conclusions are drawn in Section \ref{Sec:conclusion}.

\section{Explainable machine learning based transform and analysis}
\label{sec:dataDrivenTrCodec}

\subsection{Problem Formulation}
\label{sec:dataDrivenTr}

Transform in image/video compression is general linear and aims to improve transform performances, such as energy compaction and decorrelation, for the transformed coefficients. Let $ \textit {\textbf x}=\{x_i\}$ be an input source, and it is forward transformed to output $\textit{\textbf y}={\{y_k\}}$ in another transform domain as

\begin{equation}\label{eq:transformFWD}
{y_k} = \sum\limits_{i = 0}^{K-1} {{x_i}{a_{k,i}}} \ \ or\ \ \textit{\textbf y}= {\textbf A}\textit {\textbf x},
\end{equation}
where $\textit {a}_{k,i}$ is the transform element in the forward transform kernel ${\textbf A}$. The inverse transform from $\textit{y}_k$ to ${x_i}$ is presented as

\begin{equation}\label{eq:transformINV}
{x_i} = \sum\limits_{k = 0}^{K-1} {{y_k}{u_{k,i}}} \ \ or\ \ \textit{\textbf x}= {\textbf U}\textit {\textbf y},
\end{equation}
where $\textit {u}_{k,i}$ is the transform element in the inverse transform matrix ${\textbf U}$. ${\textbf U}$ is an inverse matrix of ${\textbf A}$ satisfying ${\textbf U}={\textbf A}^{-1}$ and ${\textbf {UA}}={\textbf {AU}}={\textbf I}$, where ${\textbf I}$ is the identity matrix. If the transform is orthogonal, which means the rows of transform matrix are an orthogonal basis set, the inverse transform matrix ${\textbf U}$ satisfies ${\textbf U} = {\textbf A}^{-1}={\textbf A}^{T}$.

As a machine learning based transform, ${\textbf A}$ is estimated from data samples ${\cal D} = [{d_0},...,{d_{T - 1}}]$. Generally, transforms are learned from subspaces of data samples with different learning strategies. Given a transform set ${\cal A}$, the optimal transform ${\textbf A}^*$ is selected from ${\cal A}$ through solving the optimization problem expressed as
\begin{equation}\label{eq:objectiveFunA}
{\textbf A}^* = \mathop {\arg \max }\limits_{{{\textbf A}_i} \in {\textbf \cal A}} M( \textit {\textbf y}),
\end{equation}
where $M(\textit {\textbf y})$ indicates a target transform performance of the transformed coefficient $ \textit {\textbf y}$. The optimal transform could be selected by maximizing the value of $M(\textit {\textbf y})$. In video coding, $M$(\textit {\textbf y}) can be defined as but not limited to the energy compaction or decorrelation capabilities, which relates to the compression efficiency. For example, DCT is predefined as a general orthogonal transform for all block residuals. KLT kernel is derived by maximizing the decorrelation, which varies for each block residual. Although KLT outperforms DCT on energy compaction and decorrelation, KLT based video coding requires to transmit kernel information for each block which causes large overhead bits. Saab transform is able to learn statistics for a large number of blocks, which is a possible solution for improving the coding performance of existing codecs.

\subsection{Saab Transform }
\label{sec:Saab}

\begin{figure}[!t]
  \centering
  \includegraphics[width=0.5\textwidth]{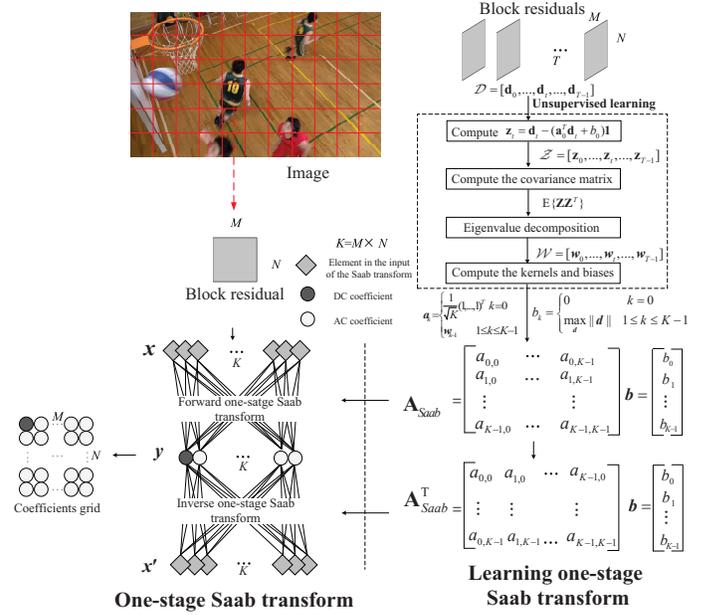}
  \caption{ Diagram of learning and testing of an one-stage Saab transform.}\label{fig:Saabtransform}
\end{figure}

Saab transform \cite{Kuo2019} is conducted as a data-dependent, multi-stage and nonseparable transform in a local window to get a local spectral vector. Saab transform is explainable machine learning based transform which is algorithmic transparency, human comprehensible, more robust to perturbations and can be improved layer by layer. Diagram of testing and training the 2D one-stage Saab transform is presented in Fig. \ref{fig:Saabtransform}, where the left part is Saab transform and the right part is learning transform kernels.

Given an $M\times N$ dimensional input $\textbf{\textit{x}}$ in the space $\textbf{R}^{M \times N}$, which is rearranged into a vector in lexicographic order as
\begin{equation}\label{eq:arrayX}
\begin{aligned}
{\textit{\textbf{x}}} = [{x_{00}},{x_{01}},...,{x_{0,N - 1}},{x_{1,0}},{x_{1,1}},...,{x_{1,N - 1}} \\
\quad \quad \quad \;\;\;...,{x_{M - 1,0}},...,{x_{M - 1,N - 1}}]^T.
\end{aligned}
\end{equation}
Then, output transformed coefficients from Saab transform can be computed as
\begin{equation}\label{eq:arrayY}
{y_k} = \sum\limits_{j = 0}^{K - 1} {{a_{k,j}}{x_j} + {b_k}}  = {\bf{a}}_k^T{\textit{\textbf{x}}} + {b_k},
\end{equation}
where ${\bf{a}}_k$ are transform kernels and ${\bf{b}}_k$ is the bias, $K$=$M\times N$, $k = 0,1,...,K - 1$. In Saab transform, DC kernel and AC kernels, are composed of $\textbf{A}_{Saab}$=$\{ {{\bf{a}}_k}\} _0^{K - 1}$ and $\textbf{\textit{b}}$=$\{ {{\bf{b}}_k}\} _0^{K - 1}$, which are unsupervised learned from the training dataset ${\cal D} = [{\textbf{d}_0},...,{\textbf{d}_{T - 1}}]$, as illustrated at the right part of Fig. \ref{fig:Saabtransform}. The number of samples in the training dataset, i.e., $T$, is around 60K. $\textit{\textbf{y}}$ is generally organized as a coefficients grid. In the process of the forward Saab transform, for input $\textbf{\textit{x}}$ in the space $\textbf{R}^{M \times N}$, the DC and AC coefficients for $\textit{\textbf{y}}$ are computed separately as
 \begin{itemize}
\item \emph{DC Coefficient}: The DC coefficient is computed with ${y_0}=\frac{1}{{\sqrt K }} \sum\limits_{j = 0}^{K - 1} {x_j}+{b_0}$, where DC kernel ${{\bf{a}}_0} = \frac{1}{{\sqrt K }}{(1,...,1)^T}$ and the corresponding bias $b_0$=0.
\item \emph{AC Coefficients}: Firstly, ${\bf{z}}{'}$ is computed as ${{\bf{z}}{'}} = {\bf{x}} - ({\bf{a}}_0^T{\bf{x}} + {b_0}){\bf{1}}$, where ${\bf 1}={\bf c}/||{\bf c}||$, and ${\bf c}=(1, 1, \cdots, 1, 1)$
is the constant unit vector. Then, AC coefficient is computed as ${y_k} = \sum\limits_{j = 0}^{K - 1} {{a_{k,j}}{z{'}_j} + {b_k}}  = {\bf{a}}_k^T{\textit{\textbf{z}}{'}} + {b_k}.$  AC kernel ${\bf{a}}_k$ is the eigenvectors $\textbf{\textit{w}}_k$ of the covariance matrix ${\bf{C}} = E\{ {\bf{Z}}{\bf{Z}}^T\}$, where ${\cal Z} = [{\textbf{z}_0},...,{\textbf{z}_t},...,{\textbf{z}_{T - 1}}]$ and ${{\bf{z}}} = {\bf{d}} - ({\bf{a}}_0^T{\bf{d}} + {b_0}){\bf{1}}$ are derived from the training dataset ${\cal D}$. The corresponding bias is ${b}_k=\mathop {\max }\limits_\textbf{d} ||\textbf{d}||$.
\end{itemize}

The inverse Saab transform process is symmetric to its forward Saab transform. Since one-stage Saab transform is orthogonal \cite{Kuo2019}, the inverse Saab transform ${\textbf U_{Saab}}$ is simply the transpose of the forward Saab transform, noted as ${\textbf U}_{Saab}={\textbf A}_{Saab}^T$. The vector $\textbf{\textit{y}}$ is inverse transformed into $\textbf{\textit{x}}'$ with ${x_k}' = {{\bf{a}}_k}'(\{ y_k\} _1^{K - 1} - \{ b_k\} _1^{K - 1}) + {y_0}$, where ${{\bf{a}}_k}'$=$[{a_{1,k}}$, ${a_{2,k}}$, $...$, ${a_{K-1,k}}]$, $k = 0,1,...,K - 1$.

Multi-stage Saab transform \cite{Kuo2019} can be built by cascading multiple one-stage Saab transforms, which can be used to extract high-level recognition features. To solve the sign confusion problem for pattern recognition, the input of the next stage is shifted to be positive by the bias. It has been exploited in the handwritten digits recognition and object classification \cite{Kuo2019,Chen2020}. In this paper, we explore the potential of Saab transform for video coding.

\subsection{ Energy Compaction and Decorrelation Capabilities of Saab Transform}
\label{sec:Decorrelation}

\begin{figure}[!t]
  \centering
  \includegraphics[width=0.51\textwidth]{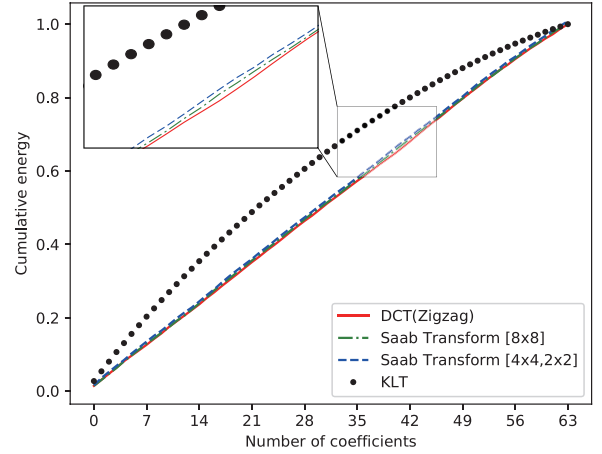}
  \caption{ Energy compaction $E( \textit {\textbf y})$ comparison among KLT, DCT and Saab transforms.}
  \label{fig:energycompaction}
\end{figure}

In video data compression, one discipline of transform is to save bits by transforming input $\textbf{\textit{x}}$ to another domain with fewer non-zero elements, which is noted as energy compaction. Therefore, the energy compaction is mathematically defined as \cite{Lohscheller1984}
\begin{equation}\label{eq:EnergyCompaction}
E(\textit{\textbf y})=\frac{{\sum\limits_{k = 0}^i {y_k^2} }}{{\sigma_\textbf{\textit{x}}^2}},
\end{equation}
where $\sigma _\textbf{\textit{x}}^2$ is the variance of the input $\textbf{\textit{x}}$ and $i$ is the number of coefficients. Accordingly, we analyze the energy compaction of 8$\times$8 transforms for video coding, i.e., KLT, DCT, one-stage Saab transform and two-stage Saab transform.
Without losing generalization, both the one-stage (denoted as ``Saab Transform [8$\times$8]") and the two-stage Saab transform (denoted as ``Saab Transform [4$\times$4,2$\times$2]") were learned from over 70K 8$\times$8 luminance (Y) block residuals of ``Planar" mode collected from encoding the video sequence ``FourPeople" with Quantization Parameters (QPs) $\in$ $\{$22, 27, 32, 37$\}$ in HEVC. Only one Saab transform was trained off-line and applied to transform all blocks in ``Saab Transform [8$\times$8]" and ``Saab Transform [4$\times$4,2$\times$2]", respectively. Then, other 500 of 8$\times$8 luminance (Y) block residuals were randomly selected to compute the energy compaction of KLT, DCT and these two Saab transforms. Fig. \ref{fig:energycompaction} shows the energy compaction $E( \textit {\textbf y})$ comparison among KLT, DCT, the one-stage Saab transform and two-stage Saab transform, and we can have the following two key observations: 1) KLT outperforms DCT by a large margin on the energy compaction, since KLT is specified for each block and
DCT is fixed and pre-defined for all blocks.
DCT is desired to obtain the best energy compaction performance for the source following Gaussion. In many cases, the assumption of the Gaussion source is not always satisfied in video coding of various contents and settings. For video content with complex texture which is not exact Gaussian source, the energy compaction performance gap between DCT and the best KLT is larger than the content with smooth texture which is closer to be Gaussian source.  2) One-stage and two-stage 8$\times$8 Saab transforms, which learn the fixed transform kernels off-line from training data and then are applied to transform all testing blocks, perform better than DCT in energy compaction. Therefore, 8$\times$8 Saab transform has the potential to improve the coding performance of the existing video codecs with the single choice of DCT.

Another discipline of transform in video coding is removing redundancy or correlation of the input signals $\textit{\textbf x}$ via transformation, i.e., decorrelation. To evaluate the decorrelation capability of a transform, we measure the decorrelation cost of transformed coefficients $\textit{\textbf y}$ with its covariance as
\begin{equation}\label{eq:DecorrelationFun}
\begin{array}{l}
C(\textit{\textbf y}) = \sum\limits_{i \ne j} {|cov({y_i},{y_j})|} \\
\quad \quad \; = \sum\limits_{i \ne j} {|E\{({y_i} - {\mu _i})({y_j} - {\mu _j})\}|} ,\;0 \le i,j \le K - 1
\end{array},
\end{equation}
where $cov$($y_i$,$y_j$) is the covariance between $y_i$ and $y_j$, $i \ne j$.  $\mu_i$ and $\mu_j$ are the mean of $y_i$ and $y_j$. Smaller $C(\textit{\textbf y})$ value indicates a better decorrelation capability of a transform. The value of $C(\textit {\textbf y})$ in the transform domain of KLT is 0, which means $y_i$ and $y_j$ are completely independent and their correlation is 0, as $i \ne j$. In other words, redundancy is minimized as 0 among the elements $y_i$ in the transformed coefficients.

\begin{table}[!t]
\begin{center}
\caption{Decorrelation capability $C( \textit {\textbf y})$ comparison among KLT, DCT and Saab transforms.}\label{tab:TransformGain}
\begin{tabular}{|c|c|c|c|c|c|} \hline
 \multirow{2}{*}{ Sequence }  & \multirow{2}{*}{ QP } &  \multicolumn{4}{c|}{Decorrelation cost $C( \textit {\textbf y})$ }\\
 \cline{3-6}
  &  & \multirow{2}{*}{ DCT } & \multicolumn{2}{c|}{Saab Transform } & \multirow{2}{*}{ KLT }\\ \cline{4-5}
  &  &   & [8$\times$8] & [4$\times$4,2$\times$2] &\\ \hline
 \multirow{4}{*}{BasketballDrill} &  22 & 581.78 & 574.26 & 582.46 & \multirow{13}{*}{0}\\ \cline{2-5}
  &  27 & 1453.53 & 1309.17 & 1357.91 & \\  \cline{2-5}
  &  32 & 3266.85 & 2691.37 & 2769.59 & \\  \cline{2-5}
  &  37 & 5886.47 & 4574.13 & 4802.21 & \\  \cline{0-4}
 \multirow{4}{*}{RaceHorses} &  22 & 588.19 & 586.50 & 591.71 & \\ \cline{2-5}
  &  27 & 1385.87 & 1361.44 & 1382.41 & \\  \cline{2-5}
  &  32 & 3911.97 & 3814.71 & 3887.13 & \\  \cline{2-5}
  &  37 & 8281.33 & 7818.34 & 8148.87 & \\  \cline{0-4}
 \multirow{4}{*}{FourPeople} &  22 & 358.67 & 371.54 & 381.89 & \\ \cline{2-5}
  &  27 & 1042.21 & 1010.10 & 1054.60 & \\  \cline{2-5}
  &  32 & 1348.89 & 1331.15 & 1394.65 & \\  \cline{2-5}
  &  37 & 2842.04 & 2691.33 & 2950.05 & \\  \cline{0-4}
 \multicolumn{2}{|c|}{Average } & 2578.98 & 2344.50 & 2441.96 & \\ \hline
\end{tabular}
\end{center}
\end{table}

Experimental analyses on the decorrelation capability of one-stage Saab transform, two-stage Saab transform and DCT for 8$\times$8 block residuals were performed. Saab transform kernels were learned from three video sequences in $\{$``BasketballDrill", ``RaceHorses", ``FourPeople"$\}$. For each video sequence, the value of $C( \textit {\textbf y})$ was computed for 500 of block coefficients randomly selected in the transform domain of each transform. As to elements in the block residual are either negative integer or positive integer, the expectation of the block residual element is supposed to be zero mean, i.e., $\mu_i = \mu_j=0$, in computing the decorrelation cost defined as $C(\textit{\textbf y})$. In Table \ref{tab:TransformGain}, decorrelation cost in terms of Eq. \ref{tab:TransformGain} of KLT, one-stage Saab transform, two-stage Saab transform and DCT are 0, 2344.50, 2441.96 and 2578.98 on average. We can have the following three key observations: 1) the decorrelation cost $C(\textit{\textbf y})$ of KLT is 0, which is confirmed to be the best. 2) Saab transform performs better than DCT on average. For smaller QP, i.e., 22, as well as for some specific video sequences, e.g., ``FourPeople", DCT performs better on the decorrelation than two-stage Saab transform, which presents the well design of DCT in some cases but not all of them. 3) One-stage Saab transform is better than two-stage Saab transform on decorrelation, mainly because the increase of stages in Saab transform is designed to be beneficial for distinguishing the category of blocks, but not necessary for the decorrelation $C(\textit{\textbf y})$. This motivates us to explore one-stage 8$\times$8 Saab transform based video coding to improve the coding efficiency.

\section{Saab transform based intra video coding}
\label{sec:frmwork}

In this section, Saab transform is applied to video codec to improve intra video coding efficiency. Firstly, a framework of Saab transform based intra video coding is presented. Then, intra mode dependent Saab transform is developed. Finally, three integration strategies of Saab transform for intra video coding are proposed to improve coding performances.

\subsection{Framework of Saab Transform based Intra Video Coding}
\label{sec:frmworkInstance}

\begin{figure}[!t]
  \centering
  \includegraphics[width=0.51\textwidth]{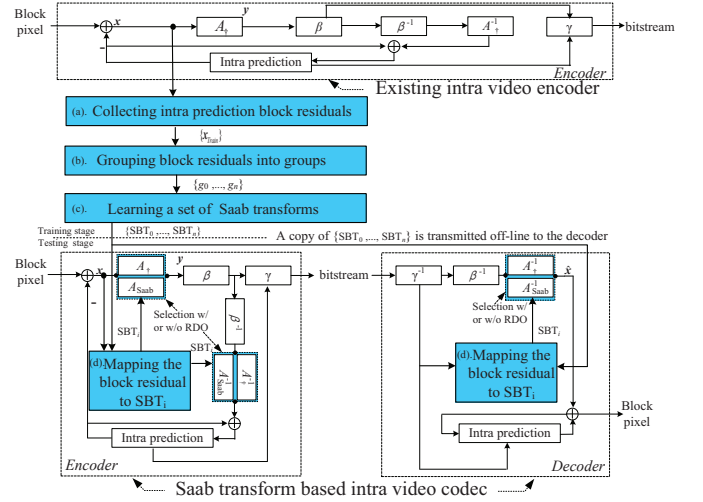}
  \caption{ Framework of Saab transform based intra video coding.}
  \label{fig:trFramework}
\end{figure}

Fig. \ref{fig:trFramework} illustrates the proposed coding framework of Saab transform based intra video coding, where $A_{Saab}$ denotes Saab transform, $\beta $ denotes quantizer and $\gamma$ denotes entropy coding. The Saab transform contains the learning kernel stage and transform stage, which consists of four key components, including (a) collecting intra prediction block residuals, (b) dividing block residuals into groups based on intra mode, (c) learning off-line a set of intra mode dependent Saab transforms and (d) selecting kernel based on the best intra prediction mode to do the forward and inverse Saab transforms.

At the stage of learning Saab transform kernels, the intra prediction block residuals ${\cal D }$=$\{\textbf{\emph{x}}_{Train}\}$ are collected off-line from conventional DCT based video encoder. Since the distribution of the residual data highly depends to the intra mode \cite{Zhao2019}, all intra modes are divided into $n$ mode sets, $\textbf{M}_i$, $i\in [0,n]$ and $n<=35$ for HEVC. Then, block residuals ${\cal D }$=$\{\textbf{\emph{x}}_{Train}\}$ are divided into groups $\{g_i\}$ regarding their intra mode whether in $\textbf{M}_i$ or not. A number of Saab transform kernels $\{$SBT$_0$,...,SBT$_n\}$ are learned individually based on the intra mode in $\textbf{M}_i$ and their block residual groups $\{g_i\}$. n is 23 here in the proposed scheme. For example, only blocks in $g_i$ are used to train SBT$_i$. Note that this is an off-line training process that various video sequences and settings can be used to train the Saab transform kernels. The complexity of the Saab transform training is negligible to the codec if it is off-line. Also, the trained Saab transform kernels are transmitted only once and stored at client side for inverse transform in decoding.

At the stage of using the Saab transform kernels, the learned Saab transform kernels $\{$SBT$_0,...,$SBT$_n\}$ are utilized to transform block residuals based on the intra mode, regarding to the learning schemes in section \ref{sec:IntraPredGrouping} different from the existing mode dependent transform selection strategy, e.g., MDDT\cite{Arrufat2014}. For example, SBT$_i$ will be used to transform the block residuals from mode in $\textbf{M}_i$. Note that there are several cases to integrate Saab transform into the video encoder. One is to replace the conventional DCT with the Saab transform. The other is to add the Saab transform as an alternative transform option and select the optimal one between Saab transform $A_{Saab}$ and the conventional transform $A_\dag$, i.e., DCT, by RD cost comparison. In the latter case, an signaling flag of choosing Saab transform or DCT is required to be encoded and transmitted to the client side for decoding.

At the decoder side, if the conventional DCT is replaced with Saab transform, based on the intra mode in $\textbf{M}_i$, the Saab transform kernel SBT$_i$ will be used in the inverse Saab transform to reconstruct block residuals. Otherwise, based on the signalling flag and intra mode in $\textbf{M}_i$, either DCT or SBT$_i$ will be selectively used in the inverse transform.

\begin{figure}[!t]
  \centering
  \includegraphics[width=0.4\textwidth]{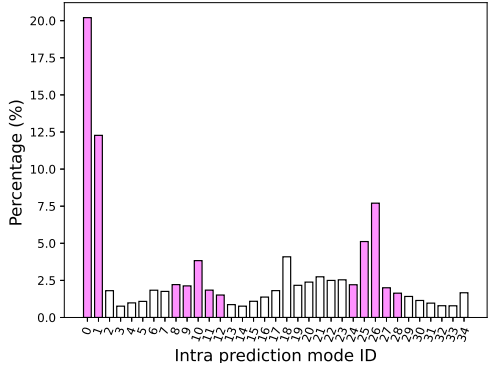}
  \caption{Percentage of luminance (Y) blocks coded with each intra prediction mode indexed by 0 $\sim$ 34.}\label{fig:IntraPreditionModePercentage}
\end{figure}

\begin{figure}[!t]
  \centering
  \includegraphics[width=0.35\textwidth]{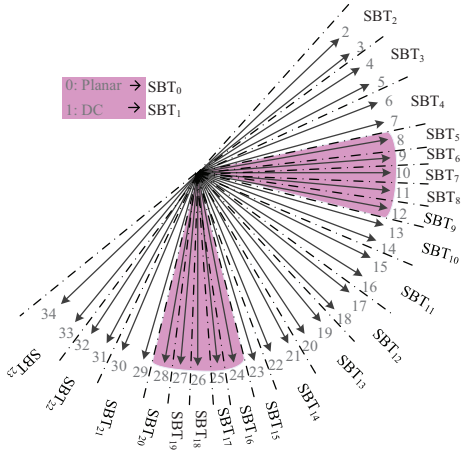}
  \caption{Grouping intra prediction block residuals on the basis of intra prediction modes. Saab transforms, noted as SBT$_k$, 1 $\le k \le$  24, are applied to the corresponding block residuals.}\label{fig:intraPredSTK}
\end{figure}

\subsection{Intra Mode Dependent Saab Transform}
\label{sec:IntraPredGrouping}

Saab transform is a data-driven transform which is learned based on the statistical characteristics of source input. However, the statistical characteristics of intra prediction block residuals $\{\textit {\textbf x}\}$ are depending on intra prediction accuracy \cite{Cai2015} as well as image texture \cite{Xu2007}. There is still significant directional information left in the intra prediction block residuals . The intra prediction block residuals generated by the same intra prediction mode still have highly varying statistics. Therefore, it is necessary to learn Saab transforms based on the statistical characteristics of the angular intra mode and develop intra mode dependent Saab transform.

We propose to divide intra prediction block residuals into groups ${g_i}$ in terms of intra prediction modes $\textbf{M}_i$, which are then used to learn Saab transform kernel SBT$_i$. The statistical characteristics of block residuals from single intra prediction mode is relatively easier to be represented, in comparison to the case of multiple intra prediction modes. On the other hand, Saab transform learned unsupervisedly for block residuals from single intra prediction mode, noted as Fine-Grained Saab Learning (FGSL), may have better performance than those learned for multiple intra prediction modes, noted as Coarse-Grained Saab Learning (CGSL). However, FGSL trains SBT$_i$ for each intra mode, $0\le i\le 35$. It means there are 35 SBTs for each Transform Unit (TU) size in HEVC, and even more kernels shall be learned for standards beyond HEVC, which significantly increases the difficulties in codec design. In addition, the ratio of blocks for each mode distributes unevenly. The distributions of 35 intra prediction modes were statistically analyzed on the number of 8$\times$8 luminance (Y) block residuals encoded by each of these modes. 100 frames of each video sequence in $\{$``BasketballDrill", ``FourPeople", ``RaceHorses"$\}$ were encoded with four QPs in $\{$22, 27, 32, 37$\}$ in HEVC. As shown in Fig. \ref{fig:IntraPreditionModePercentage}, block residuals of ``Planar", ``DC", ``Horizontal" and ``Vertical" and their neighboring modes have more percentages than those of rest modes, which indicate these modes have higher impacts on the coding efficiency than the others. Therefore, considering the coding efficiency and complexities of designing the codec, we will train Saab transform kernels for ``Planar (0)", ``DC (1)", ``Horizontal (10)" and ``Vertical (26)" and their neighboring modes (8 $\sim$ 12 and 24 $\sim$ 28), shown as in Category B in Table \ref{tab:trHVSetSimilarity} with FGSL scheme. Saab transform kernels for the rest modes, shown as in Category A in Table \ref{tab:trHVSetSimilarity}, are trained with CGSL scheme. We considered the coding efficiency and complexity of designing the codec. 24 Saab transforms for intra predicted block residuals are trained with FGSL and CGSL schemes. In comparison to the other KLT derivation based intra prediction mode dependent transform selection which design adapted transform to each of the 35 intra prediction modes, FGSL and CGSL could save more memory to store the transform kernel as well as more complexity of designing the transform kernels. Whereas to be compared with the DCT/DST transform based intra prediction mode dependent multiple transforms, FGSL and CGSL is supposed to learn the statistics better for block residuals through grouping block residuals in terms of intra prediction modes. The details of the FGSL and CGSL schemes are:

\begin{table}[!t]
	\small
	\setlength{\abovecaptionskip}{0.cm}
	\setlength{\belowcaptionskip}{-0.cm}	
	\caption{Two training strategies for mode dependent Saab transform. }\label{tab:trHVSetSimilarity}
	\begin{center}
        \begin{tabular}{ c | c |c  }
			\hline

              Category & Mode ID &Saab Learning Scheme\\
              \hline
              \hline
              A        & 2 $\sim$ 7, 13 $\sim$ 23 , 29 $\sim$ 34&CGSL\\
              \hline
              \rowcolor{mygray}
              B       & 0 $\sim$ 1, 8 $\sim$ 12, 24 $\sim$ 28 &FGSL\\
              \hline
		\end{tabular}
	\end{center}	
\end{table}

\begin{itemize}
\item \emph{CGSL scheme for modes in Category A}: Grouping the intra prediction block residuals generated by intra prediction mode indexed by mode ID $i$ with block residuals related to intra prediction mode ID $i-1$ to learn SBT$_k$. SBT$_k$ will be applied to the block residuals of intra prediction mode with mode ID $i$.

\item \emph{FGSL scheme for modes in Category B}: Grouping the intra prediction block residuals generated by intra prediction mode indexed by mode ID $i$ ($i \ge 3$) with block residuals related to the intra prediction modes indexed by mode ID $i-1$ to learn SBT$_k$. Different from Scheme CGSL, the learned Saab transform SBT$_k$ will be applied to block residuals generated by intra prediction modes indexed by mode ID $i-1$ and $i$. Residuals generated by ``Planar" (indexed with $i=0$) and ``DC" (indexed with $i=1$) are grouped respectively as two groups for learning and testing the SBT$_k$.
\end{itemize}

\begin{table*}[t]
	\scriptsize
	\begin{center}
   \begin{threeparttable}[b]	
	\caption{Intra mode dependent Saab transform set $\{$SBT$_k$$\}$ and integration strategies. }\label{tab:trSet}
    \setlength{\tabcolsep}{1.5mm}{
		\begin{tabular}{>{\centering \arraybackslash }m{53pt}  |>{\centering \arraybackslash }m{38pt} |>{\centering \arraybackslash }m{2pt}  >{\centering \arraybackslash }m{2pt}   >{\centering  }m{2pt}  >{\centering \arraybackslash }m{2pt}  >{\centering  }m{2pt}  >{\centering  }m{2pt}   >{\centering \arraybackslash }m{2pt}  >{\centering \arraybackslash }m{2pt}   >{\centering \arraybackslash }m{2pt}  >{\centering \arraybackslash }m{2pt}  >{\centering \arraybackslash }m{2pt}  >{\centering \arraybackslash }m{2pt}  >{\centering \arraybackslash }m{2pt} >{\centering \arraybackslash }m{2pt} > {\centering \arraybackslash }m{2pt} >{\centering \arraybackslash }m{2pt}  >{\centering \arraybackslash }m{2pt}   >{\centering  }m{2pt}  >{\centering \arraybackslash }m{2pt}  >{\centering  }m{2pt}  >{\centering  }m{2pt}   >{\centering \arraybackslash }m{2pt}  >{\centering \arraybackslash }m{2pt}   >{\centering \arraybackslash }m{2pt}  >{\centering \arraybackslash }m{2pt}  >{\centering \arraybackslash }m{2pt}  >{\centering \arraybackslash }m{2pt}  >{\centering \arraybackslash }m{2pt} >{\centering \arraybackslash }m{2pt} > {\centering \arraybackslash }m{2pt} >{\centering \arraybackslash }m{2pt}  >{\centering \arraybackslash }m{2pt}  >{\centering \arraybackslash }m{2pt}  >{\centering \arraybackslash }m{2pt} >{\centering \arraybackslash }m{2pt} > {\centering \arraybackslash }m{2pt}  }
			\hline
             \multirow{2}{53pt}{Integration Strategies}&\multirow{2}{*}{  Transform }& \multicolumn{35}{c}{Intra Mode ID}\\
             \cline{3-37}
               & &0 & 1 & 2 & 3 & 4 & 5 & 6 & 7 & 8 & 9 & 10 & 11 & 12 & 13 & 14 & 15 & 16 & 17 & 18 & 19 & 20 & 21 & 22 & 23 & 24 & 25 & 26 & 27 & 28 & 29 & 30 & 31 & 32 & 33 & 34 \\
             \hline
             \hline
             \multirow{1}{*}{  s$_I$ \tnote{1} }&SBT / DCT& 0  & 1 & 2 & 2 & 3 & 3 & 4 & 4 & \multicolumn{5}{|c|}{ DCT } & 10 & 10 & 11 & 11 & 12 & 12 & 13 & 13 & 14 & 14 & 15 &  \multicolumn{5}{|c|}{ DCT } & 21 & 21 & 22 & 22 & 23  & 23 \\
             \hline
             \hline
              \multirow{2}{*}{  s$_{II}$ \tnote{2} } & SBT index & 0 & 1 & 2 & 2 & 3 & 3 & 4 & 4 & \multicolumn{5}{|c|}{ N/A } & 10 & 10 & 11 & 11 & 12 & 12 & 13 & 13 & 14 & 14 & 15 & \multicolumn{5}{|c|}{ N/A } & 21 & 21 & 22 & 22 & 23  & 23 \\\cline{2-37}
              & DCT &  \multicolumn{35}{c}{  DCT }\\
              \hline
              \hline
              \multirow{2}{*}{  s$_{III}$ \tnote{2} }  & SBT index   & 0  & 1 & 2 & 2 & 3 & 3 & 4 & 4 & 5 & 6 & 7 & 8 & 9 & 10 & 10 & 11 & 11 & 12 & 12 & 13 & 13 & 14 & 14 & 15 & 16 & 17 & 18 & 19 & 20 & 21 & 21 & 22 & 22 & 23  & 23 \\\cline{2-37}
              & DCT   & \multicolumn{35}{c}{ DCT } \\
             \hline
             \hline
		\end{tabular}}
       \begin{tablenotes}
         \item[1] Either DCT or SBT$_k$ is used depending on the intra mode.
         \item[2] The optimal transform is selected from DCT and SBT$_k$ with RDO. One bit signalling flag is transmitted to indicate the type, where 0 / 1 indicate DCT / SBT$_k$.
       \end{tablenotes}
   \end{threeparttable}
   	\end{center}	
\end{table*}

Results of the proposed grouping scheme are illustrated in Fig. \ref{fig:intraPredSTK}, where 24 of the Saab transform kernels are learned and applied to intra prediction block residuals, noted as SBT$_k, 0\le k\le 23$. FGSL is adopted by block residuals generated by intra prediction modes indexed by mode ID in $\{$0$\sim$7, 13$\sim$23, 29$\sim$34$\}$. According to FGSL, SBT$_0$ and SBT$_1$ are learned for block residuals of ``Planar" and ``DC" separately. SBT$_k$ with $k$ in $\{$2$\sim$4, 10$\sim$15, 21$\sim$23$\}$ are learned from block residuals generated by intra prediction modes pairs in $\{$(2,3), (4,5), (6,7), (13,14), (15,16), (17,18), (19,20), (21,22), (22,23), (29,30), (31,32), (33,34)$\}$. These Saab transforms are applied to corresponding block residuals regarding to the FGSL, except SBT$_{15}$, which is learned from block residuals corresponding to intra prediction modes pairs (22,23) but (23,24), and only applied to the block residuals generated by the intra prediction mode indexed by 23, as CGSL is utilized to learn the Saab transform for block residuals of the intra prediction mode indexed by 24. CGSL is adopted by intra prediction modes indexed by mode ID in $\{$8$\sim$12, 23$\sim$28$\}$. Regarding CGSL, SBT$_k$ with mode ID in $\{$5$\sim$9, 15$\sim$20$\}$ are learned from block residuals generated by intra prediction modes pairs in $\{$(7,8), (8,9), (9,10), (10,11), (11,12), (12,13), (23,24), (24,25), (25,26), (26,27), (27,28)$\}$. These Saab transforms are applied to the block residuals according to CGSL.

\subsection{Integration Strategies for Saab Transform}
\label{sec:IntegrationStrategies}
Since Saab transform has good performances on energy compaction and de-correlation, as analyzed in Subsection \ref{sec:Decorrelation}, we propose three integration strategies that integrate intra mode dependent Saab transform with DCT in intra video codec to improve the coding efficiency. Table \ref{tab:trSet} shows these three integration strategies, noted as $s_{I}$, $s_{II}$ and $s_{III}$, for Saab transform based intra video codec. In $s_{I}$ , each intra prediction mode adopts either Saab transform or DCT for transform coding. Intra prediction modes in $\{0\sim7, 13\sim23, 29\sim34\}$ utilize SBT$_k$ with index $k$ in $\{0\sim4, 10\sim15, 21\sim23\}$ as their transforms. For intra prediction modes that around the horizontal and vertical directions, block residuals of which retain more dynamical statistical characteristics of block content than the other modes. Replace DCT with Saab transform for these modes directly can obtain less or negative RD performance gain. Detailed analysis can be referred in Subsection \ref{sec:varOfSaabTr}. So, the original DCT in the existing intra video codec is used for these modes. To complement Saab transform with DCT, strategy $s_{II}$ in the middle row of Table \ref{tab:trSet} is derived, where SBT$_k$ is applied to intra prediction modes in $\{0\sim7, 13\sim23, 29\sim34\}$. Meanwhile, the DCT is also activated. The optimal transform is selected from DCT and SBT$_k$ by choosing the one with lower RD cost, i.e., RDO. N/A indicates there is no available SBT$_k$ for modes in $\{8\sim12, 24\sim28\}$. So, DCT is used for them directly. Furthermore, to maximize the coding efficiency, strategy $s_{III}$ is proposed, as shown in the bottom row in Table \ref{tab:trSet}. SBT$_k$, $ 0 \le k \le 23$, are applied to all 35 intra prediction modes in intra coding. Meanwhile, DCT is also activated. The optimal transform is selected from SBT$_k$ and DCT with RD cost comparison. In integration strategies $s_{II}$ and $s_{III}$, the optimal transform is selected with RDO from SBT$_k$ and DCT. In these cases, a signaling flag with 1 bit shall be added in the bitstream for each TU, where 0/1 indicates using DCT/SBT$_k$, respectively. Note that $k$ in SBT$_k$ is determined by the intra mode according to the learning scheme. The RD performance gain of Saab transform and three integration strategies in intra video coding are analyzed in following sections.

\section{RD performance and computational complexity of Saab transform based intra coding}
\label{sec:RDModelForTr}

In this section, RD performances and computational complexity of Saab transform based intra video coding are analyzed theoretically and experimentally. Firstly, we analyze the RD cost of Saab transform based intra video coding and two sufficient conditions are derived while using Saab transform to improve the RD performance. Then, these two sufficient conditions of using SBT$_k$ are validated individually with coding experiments. Finally, computational complexity of one-stage Saab transform is compared with DCT.

\subsection{Theoretical RD Cost Analysis on Saab Transform}
\label{sec:ModelDepQuant}

The objective of video coding is to minimize the distortion (D) subject to bit rate (R) is lower than a target bit rate. By introducing the Lagrangian multiplier $\lambda$, the R-D optimization problem in video coding can be formulated by minimizing RD cost ${J}(Q)$ as

\begin{equation}\label{eq:LagrangianFun}
\min {J}(Q), {J}(Q) = {D}(Q) + \lambda\cdot{R}(Q),
\end{equation}
where $Q$ is quantization step, ${D}(Q)$ and ${R}(Q)$ are distortion and bit rate at given $Q$. So, it is necessary to analyze the RD cost of Saab transform based intra coding and that of DCT based intra coding to validate its effectiveness.

\begin{figure}[!t]
\subfigure[  ]{{\includegraphics[width=0.24\textwidth]{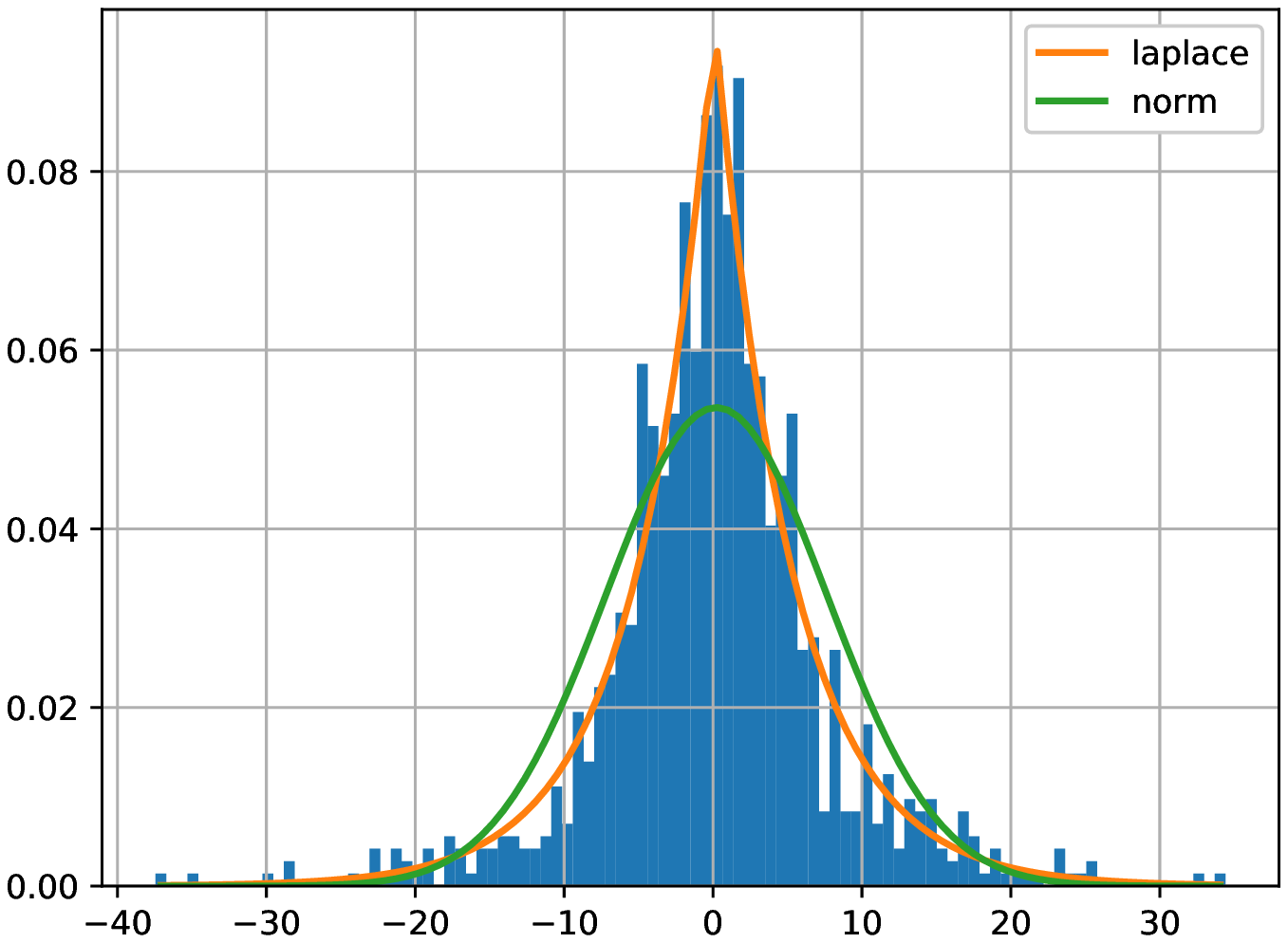}}}
\subfigure[  ]{{\includegraphics[width=0.24\textwidth]{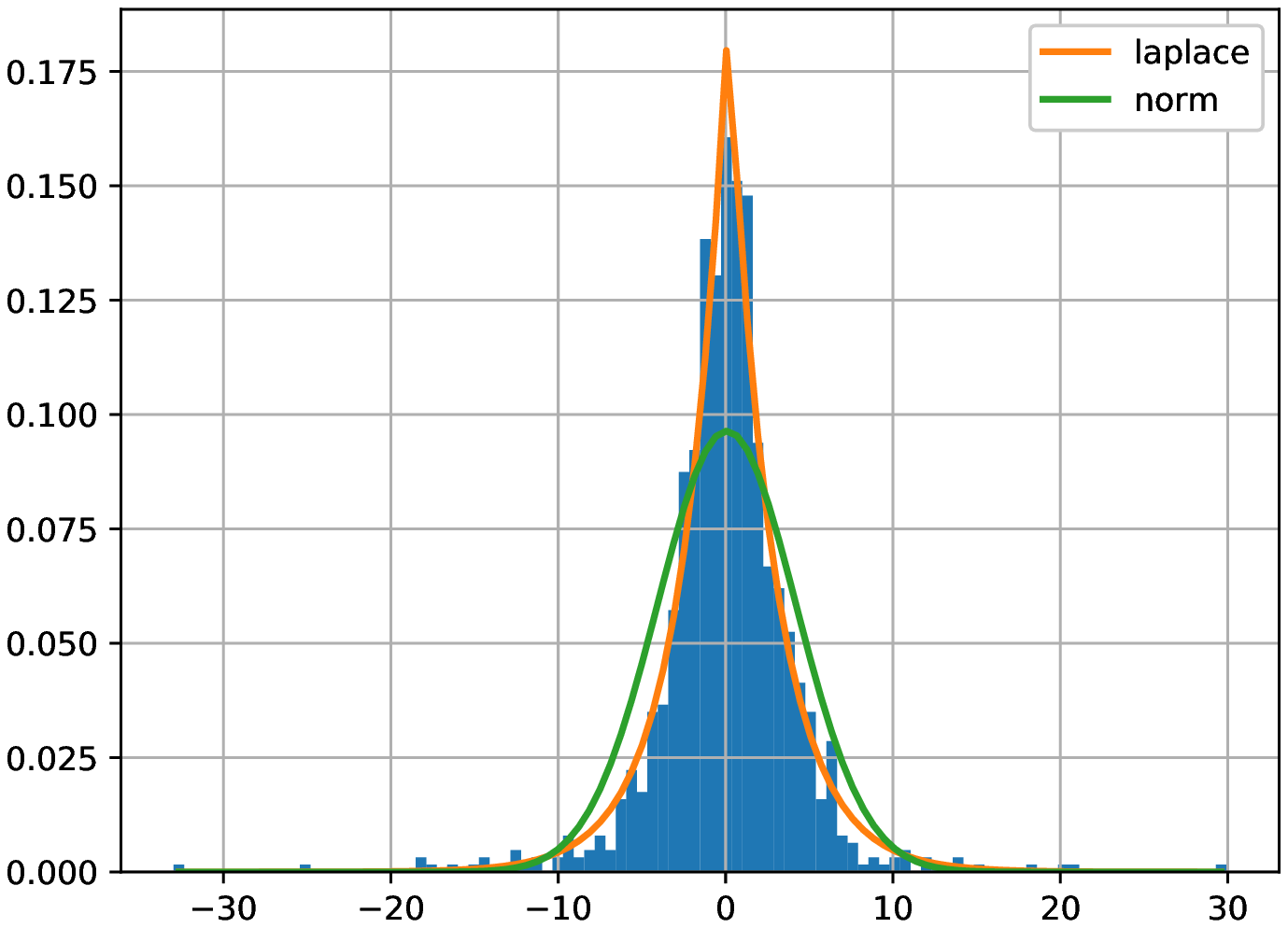}}}
\caption{Distributions of transformed coefficients via Saab transform for $8\times 8$ block residuals generated by ``Planar" mode. (a) Transformed coefficient at location (0,2) from Saab, (b) Transformed coefficient at location (5,2) from Saab. }\label{fig:Transform_Coefficients_Distribution_Saab}
\end{figure}

The rate $R({Q})$ and distortion $D({Q})$ of using the Saab transform are modelled and theoretically analyzed. The bit rate $R$ can be modelled with the entropy of the transformed coefficient $y$. Meanwhile, when the transformed coefficient $y$ is quantized with quantization step $Q$, the bit rate $R(Q)$ can be modelled as its entropy minus $\log Q$, which is \cite{Gish1968}
\begin{equation}\label{eq:GeneralRateFun}
R({Q}) \approx  - \int_{ - \infty }^{ + \infty } {{f_y}(y)\log{f_y}(y)} dy - \log {Q}.
\end{equation}
To analyze the transformed coefficient $y$ output from Saab transform, we collected 1000 of 8$\times$8 luminance (Y) intra prediction block residuals generated by ``Planar" mode from encoding ``FourPeople" in HEVC. Histograms of transformed coefficients of Saab transform at two locations of the coefficient grid, i.e., (0,2) and (5,2), are presented in Fig. \ref{fig:Transform_Coefficients_Distribution_Saab}. Saab transform coefficients for 8$\times$8 block residuals are organized as a grid of 8$\times$8 in the order of the zigzag scan. The histograms of transformed coefficients of Saab transform are approximated respectively with Laplacian distribution and Gaussian distribution, noted as ${y_{Saab}}\sim{\rm{Laplace}}(\mu_{{y_{Saab}}},\sigma _{{y_{Saab}}})$ and ${y_{Saab}}\sim{\rm{N}}(\mu_{{y_{Saab}}},\sigma _{{y_{Saab}}})$. We can observe that distributions of transformed coefficients of Saab transform generally obey Laplacian and Gaussian distributions. Meanwhile, higher accuracy can be obtained by modelling these transformed coefficients using Laplacian distribution. So, Laplacian distribution is used for modelling the transformed coefficient $y$ of Saab transform, i.e., ${f_y}(y)={{\sqrt 2 {\sigma _y}}}{e^{ - \frac{{\sqrt 2 }}{{{\sigma _y}}}|y|}}$. By applying ${f_y}(y)$ to Eq. \ref{eq:GeneralRateFun}, we can obtain the rate $R(Q)$ as

\begin{equation}\label{eq:RateFun}
\begin{array}{l}
R({Q}) \approx \log \frac{{\sqrt 2 e{\sigma _y}}}{{{Q}}}
\end{array}.
\end{equation}

Since uniform quantizer is used to quantize the transformed coefficient $y$, the range of $y$ will be partitioned into a infinite number of intervals $I_t=(q_t,q_{t+1})$. $y$ in the interval $I_t$ will be mapped to $s_t$. As $s_t$ is independent with $t$, given the quantization step size $Q=q_{t+1}-q_t$, the distortion caused by quantization $D(Q)$ can be calculated as \cite{Gray1998}
\begin{equation}\label{eq:GeneralDistFun}
D({Q}) = \sum\limits_{ - \infty }^{ + \infty } {\int_{{s_t} - 0.5}^{{s_t} + 0.5} {{{(y - {s_t})}^2}{f_y}(y)dy}},
\end{equation}
where $Q$ is the quantization step size. As the distribution of the transformed coefficient $y$ after uniform quantization still obey Laplacian, i.e., ${f_y}(y)={{\sqrt 2 {\sigma _y}}}{e^{ - \frac{{\sqrt 2 }}{{{\sigma _y}}}|y|}}$, the distortion $D(Q)$ can be approximated as \cite{Xu2007}
\begin{equation}\label{eq:DistFun}
\begin{array}{l}
D({Q}) \approx \sigma _y^2\frac{{Q^2}}{{12\sigma _y^2 + Q^2}}
\end{array}.
\end{equation}

Therefore, the RD cost of the Saab transform based intra coding scheme can be calculated as
\begin{equation}\label{eq:JLagFunG}
J \approx {\kappa _y}=\left\{ \begin{array}{l}
\sigma _y^2\frac{{Q^2}}{{12\sigma _y^2 + Q^2}} + \lambda  \cdot \log \frac{{\sqrt 2 e{\sigma _y}}}{{{Q}}}\quad {\sigma _y} > \frac{{{Q}}}{{\sqrt 2 e}}\\
\sigma _y^2\frac{{Q^2}}{{12\sigma _y^2 + Q^2}}\quad \quad \quad \quad \quad \quad \quad \; \;  {\sigma _y} \le \frac{{{Q}}}{{\sqrt 2 e}}
\end{array}\right.,
\end{equation}
where the right part is defined as ${\kappa_{{y}} }$ for further illustration. When ${\sigma _y} \le \frac{{{Q}}}{{\sqrt 2 e}}$, $R({Q})=0$, so  $J\approx {\kappa _y}= \sigma _y^2\frac{{Q^2}}{{12\sigma _y^2 + Q^2}}$.

\begin{figure}[!t]
\subfigure[  ]{{\includegraphics[width=0.24\textwidth]{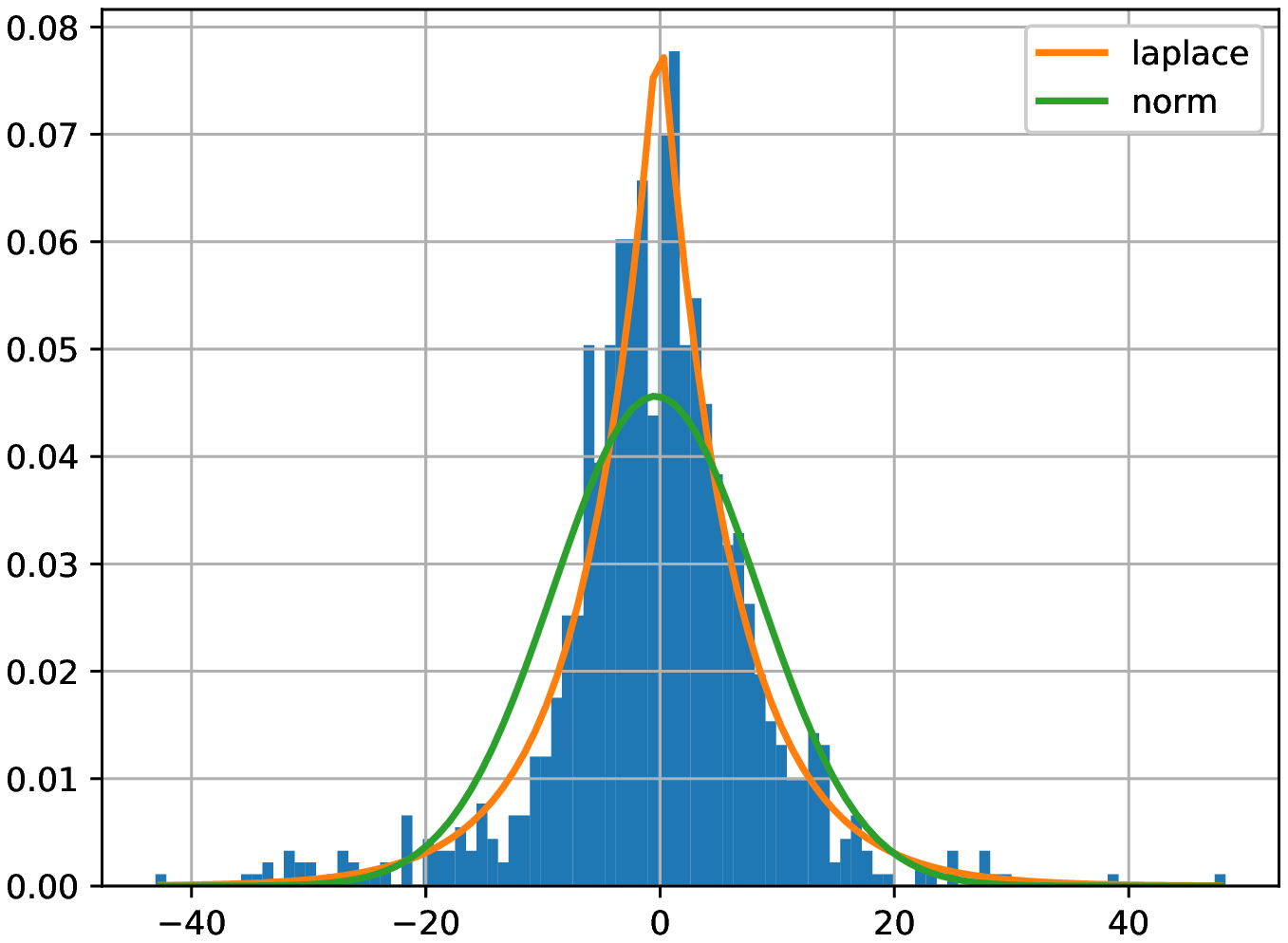}}}
\subfigure[  ]{{\includegraphics[width=0.24\textwidth]{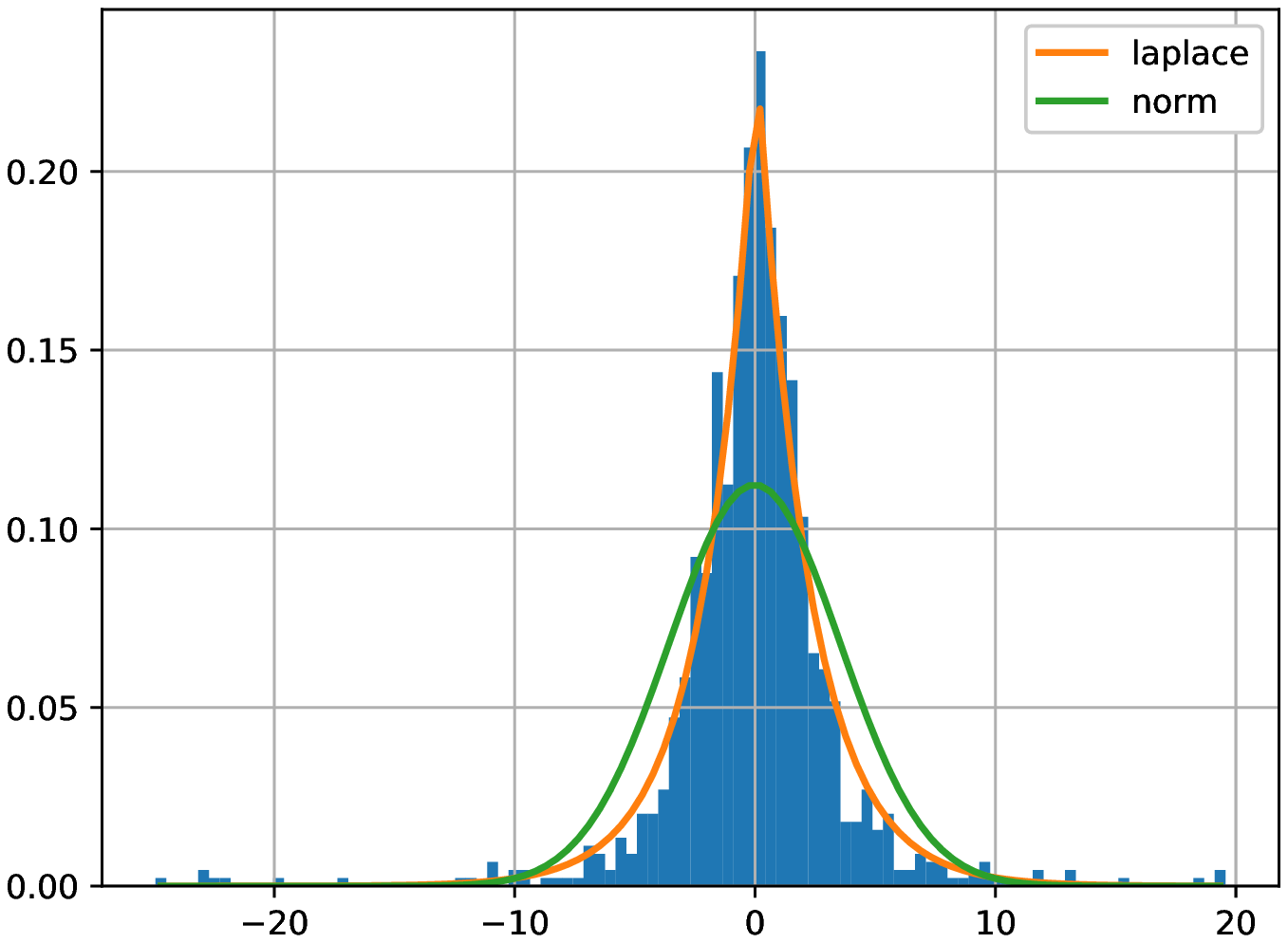}}}
\caption{Distributions of DCT transformed coefficients for $8\times8$ residual blocks generated by ``Planar" mode. (a) Transformed coefficient at location (0,2) from DCT, (b) Transformed coefficient at location (5,2) from DCT. }\label{fig:Transform_Coefficients_Distribution_DCT}
\end{figure}

Similarly, we also analyze distributions of the transformed coefficients from DCT, as shown in Fig. \ref{fig:Transform_Coefficients_Distribution_DCT}. These distributions of transform coefficients of DCT are closer to Laplacian distribution ${y_{DCT}} \sim {\rm{Laplace}}(\mu_{{y_{DCT}}} ,\sigma _{{y_{DCT}}})$ \cite{Xu2007} than Gaussian distribution ${y_{DCT}} \sim {\rm{N}}(\mu_{{y_{DCT}}} ,\sigma _{{y_{DCT}}})$. Therefore, Eq.\ref{eq:JLagFunG} can also be derived for DCT based intra coding. To differentiate Saab transform from DCT, transformed coefficients of Saab transform are noted as $y_{Saab}$ and DCT are noted as $y_{DCT}$. $\kappa_{y}$ for Saab and DCT are denoted as $\kappa_{y_{Saab}}$ and $\kappa_{y_{DCT}}$. Therefore, for block residuals, RD gain can be achieved if the transformed coefficient of Saab transform satisfies condition


\begin{equation}\label{eq:kapaIneqG}
\begin{array}{l}
{{\kappa_{y_{Saab}}}} < {{\kappa_{y_{DCT}}}}
\end{array}.
\end{equation}

Apply Eq.\ref{eq:JLagFunG} to Eq.\ref{eq:kapaIneqG}, we can obtain an inequality relates to $\sigma _{{y_{Saab}}}^2$, $\sigma _{{y_{DCT}}}^2$ and quantization step $Q$. For simplicity, we find Eq. \ref{eq:kapaIneqG} is satisfied by all quantization step $Q$ when the variances of the transformed coefficients, $ \sigma _{{y_{Saab}}}^2$ and $\sigma _{{y_{DCT}}}^2,$ satisfy condition
\begin{equation}\label{eq:varCond}
 \sigma _{{y_{Saab}}}^2 <  \sigma _{{y_{DCT}}}^2.
\end{equation}
This inequality is an sufficient but not necessary condition for Eq.\ref{eq:kapaIneqG}, which is more critical. It means transform that minimizes the output variances of transformed coefficients will improve the RD performance of a codec. Both conditions in Eq.\ref{eq:kapaIneqG} and Eq.\ref{eq:varCond} will be experimentally analyzed in detail in the following subsection so as to testify the effectiveness of Saab transform.

\begin{table*}[htb]
	\scriptsize
	
	\setlength{\abovecaptionskip}{0.cm}
	\setlength{\belowcaptionskip}{-0.cm}	
	\caption{ Comparisons between $\kappa_{y_{Saab}}$ and $\kappa_{y_{DCT}}$ for 8$\times$8 luminance (Y) block residuals from ``Planar" mode. }\label{tab:KappaForModes}
	\begin{center}
		\begin{tabular}{|>{\centering \arraybackslash }m{39pt} | >{\centering \arraybackslash }m{22pt} |  >{\centering  }m{22pt} | >{\centering \arraybackslash }m{22pt} | >{\centering  }m{22pt} | >{\centering  }m{22pt} |  >{\centering \arraybackslash }m{22pt} | >{\centering \arraybackslash }m{28pt} |  >{\centering \arraybackslash }m{28pt} | >{\centering \arraybackslash }m{26pt} | >{\centering \arraybackslash }m{28pt} | >{\centering \arraybackslash }m{28pt} | >{\centering \arraybackslash }m{22pt} | }
			\hline

            \multicolumn{1}{|c|}{QP} &  \multicolumn{3}{ c|}{22} & \multicolumn{3}{ c|}{27} & \multicolumn{3}{ c|}{32} & \multicolumn{3}{ c|}{37}\\
             \cline{1-13}
              Sequence name &  $\kappa_{y_{DCT}}$ & $\kappa_{y_{Saab}}$ & ${\Delta \kappa }$ & $\kappa_{y_{DCT}}$ & $\kappa_{y_{Saab}}$ & ${\Delta \kappa }$ & $\kappa_{y_{DCT}}$ & $\kappa_{{Saab}}$ & ${\Delta \kappa }$ & $\kappa_{y_{DCT}}$ & $\kappa_{y_{Saab}}$ & ${\Delta \kappa }$ \\
             \hline
             PeopleOnStreet &  0.1035 & 0.1036 & 0.0001 & 0.6765 & 0.6764 & -0.0001 & 4.3067 & 4.3052 & -0.0015 & 21.6978 & 21.6912 & -0.0066\\

             \hline
              RaceHorses &  0.1263 & 0.1264 & 0.0001 & 0.9002 & 0.9001 & -0.0001 & 5.3745 & 5.3743 & -0.0002 & 22.6707 & 22.6658 & -0.0049\\

             \hline
              Johnny &  0.0751 & 0.0751 & 0.0000 & 0.2819 & 0.2818 & -0.0001 & 1.3996 & 1.3982 & -0.0014 & 11.1232 & 11.1211 & -0.0021\\

             \hline
             \multicolumn{1}{|c|}{ Average } &  - & - & 0.0001 & - & - & -0.0001 & - & - & -0.0010 & - & - & -0.0045\\

             \hline
		\end{tabular}
	\end{center}	
\end{table*}
\begin{table*}[htb]
	\scriptsize
	\setlength{\abovecaptionskip}{0.cm}
	\setlength{\belowcaptionskip}{-0.cm}	
	\caption{Comparisons between $\sigma _{{y_{Saab}}}^2$ and $\sigma _{{y_{DCT}}}^2$ for 8$\times$8 block residuals when QP is 37. Four intra prediction modes ``Planar", ``DC", ``Horizontal" and ``Vertical" are tested.}\label{tab:VarianceForModes}
	\begin{center}
		\begin{tabular}{ |>{\centering \arraybackslash }m{39pt} | >{\centering \arraybackslash }m{22pt} |  >{\centering  }m{22pt} | >{\centering \arraybackslash }m{22pt} | >{\centering  }m{22pt} | >{\centering  }m{22pt} |  >{\centering \arraybackslash }m{22pt} | >{\centering \arraybackslash }m{22pt} |  >{\centering \arraybackslash }m{22pt} | >{\centering \arraybackslash }m{22pt} | >{\centering \arraybackslash }m{22pt} | >{\centering \arraybackslash }m{22pt} | >{\centering \arraybackslash }m{22pt} |>{\centering \arraybackslash }m{22pt} | }
			\hline
            \multicolumn{1}{ |c|}{Intra mode} &  \multicolumn{3}{ c|}{Planar} & \multicolumn{3}{ c|}{DC} & \multicolumn{3}{ c|}{Horizontal} & \multicolumn{3}{ c|}{Vertical}\\
             \cline{1-13}
              Sequence name &  $\sigma _{{y_{DCT}}}^2$ & $\sigma _{{y_{Saab}}}^2$ & ${\Delta \sigma^2 }$ & $\sigma _{{y_{DCT}}}^2$ & $\sigma _{{y_{DCT}}}^2$ & ${\Delta \sigma^2 }$ & $\sigma _{{y_{DCT}}}^2$ & $\sigma _{{y_{Saab}}}^2$ & ${\Delta \sigma^2 }$ & $\sigma _{{y_{DCT}}}^2$ & $\sigma _{{y_{Saab}}}^2$ & ${\Delta \sigma^2 }$ \\
             \hline
              PeopleOnStreet &  24.899 & 24.891 & -0.008 & 32.741 & 32.735 & -0.006 & 32.419 & 31.416 & -1.003 & 21.673 & 21.673 & 0.000\\
             \hline
              RaceHorses &  35.890 & 35.884 & -0.006 & 54.590 & 54.589 & -0.001 & 74.434 & 74.443 & 0.009 & 53.016 & 53.031& 0.015\\
             \hline
               Johnny &  11.908 & 11.906 & -0.002 & 18.887 & 18.797 & -0.09 & 30.122 & 30.119 & -0.003& 11.460 & 11.461 & 0.001\\
             \hline
             \multicolumn{1}{|c|}{ Average } &  - & - & -0.005 & - & - & -0.032 & - & - & -0.332 & - & - & 0.005\\
             \hline
		\end{tabular}
	\end{center}	
\end{table*}

\subsection{Experimental RD Cost Analysis on Saab Transform}
\label{sec:varOfSaabTr}

Two conditions in Eq.\ref{eq:kapaIneqG} and Eq.\ref{eq:varCond} were analyzed by comparing $\sigma_{y_{Saab}}^2$ and $\sigma_{y_{DCT}}^2$, $\kappa _{y_{{Saab}}}$ and $\kappa _{y_{{DCT}}}$ of transformed coefficients from Saab and DCT, respectively. In the experiment, coding configurations were generally the same as those in Section \ref{sec:Decorrelation}. Saab transforms were learned from 80K 8$\times$8 luminance (Y) block residuals for intra prediction modes in $\{$``Planar", ``DC", ``Horizontal", ``Vertical"$\}$. Hundreds of 8$\times$8 intra prediction block residuals from each mode were randomly collected among thousands of blocks to compute $\kappa _{y_{{Saab}}}$ and $\kappa _{y_{{DCT}}}$, where these block residuals were generated from encoding video sequences with Saab transform based intra video encoder and conventional DCT based intra video encoder with ``Planar" mode only. Three sequences with different resolutions, ``PeopleOnStreet" at 2560$\times$1600, ``Johnny" at 1280$\times$720 and ``RaceHorses" at 416$\times$240 , were tested when QP $\in$ $\{$22, 27, 32, 37$\}$. To quantify the difference between $\kappa _{y_{{Saab}}}$ and $\kappa _{y_{{DCT}}}$, ${\Delta \kappa }$ is defined as
\begin{equation}\label{eq:DeltaKappa}
{\Delta \kappa } ={\kappa _{y_{{Saab}}} - \kappa _{y_{{DCT}}}} ,
\end{equation}
where negative ${\Delta \kappa }$ indicates a better RD performance of Saab transform as compared with DCT, while positive ${\Delta \kappa }$ indicates a worse RD performance. Table \ref{tab:KappaForModes} shows $\kappa_{y_{{Saab}}}$, $\kappa_{y_{{DCT}}}$ and ${\Delta \kappa }$ for different QPs and video sequences. 
We can observe that  $\kappa_{y_{{Saab}}}$ is generally smaller than $\kappa_{y_{{DCT}}}$, and the average ${\Delta \kappa }$ are 0.0001, -0.0001, -0.0010 and -0.0045 when QP is 22, 27, 32 and 37, respectively. It means for the ``Planar" mode the Saab transform can achieve better RD performance on average when QP are 27, 32 and 37 and a little worse than DCT on RD performance when QP is 22. So, Saab transform is actually more effective than DCT on average.

In addition,  $\sigma_{y_{Saab}}^2$ and $\sigma_{y_{DCT}}^2$ are also analyzed and compared to validate the effectiveness of Saab transform. Four Saab transforms were learned from 80K 8$\times$8 luminance (Y) block residuals for intra prediction modes in $\{$``Planar", ``DC", ``Horizontal", ``Vertical"$\}$, respectively. Then, these Saab transforms were applied to block residuals of $\{$``Planar", ``DC", ``Horizontal", ``Vertical"$\}$ correspondingly. As a comparison, the same set of block residuals were also transformed by DCT. Then, $\sigma _{y_{Saab}}^2$ and  $\sigma _{y_{DCT}}^2$ were computed from the transformed coefficients of Saab transform and DCT. Four intra modes $\{$``Planar", ``DC", ``Horizontal", ``Vertical"$\}$ and three video sequences $\{$``PeopleOnStreet", ``RaceHorses", ``Johnny"$\}$ were tested. QP was fixed as 37.
The difference between  $\sigma _{y_{Saab}}^2$ and  $\sigma _{y_{DCT}}^2$, i.e., ${\Delta \sigma^2 }$, is defined as
\begin{equation}\label{eq:DeltaSigma}
{\Delta \sigma^2}  ={\sigma _{y_{Saab}}^2 - \sigma _{y_{DCT}}^2} ,
\end{equation}
where negative  ${\Delta \sigma^2 }$ indicates a better RD performance of Saab transform and positive ${\Delta \sigma^2 }$ indicates a worse RD performance as compared with DCT. $\sigma _{y_{Saab}}^2$ and  $\sigma _{y_{DCT}}^2$ are variances of transform coefficients of Saab transform and DCT, respectively. Table \ref{tab:VarianceForModes} shows $\sigma _{y_{Saab}}^2$, $\sigma _{y_{DCT}}^2$ and ${\Delta \sigma ^2 }$ for four different intra modes. We can observe that the average ${\Delta \sigma ^2 }$ of intra prediction modes ``Planar", ``DC", ``Horizontal" and ``Vertical" are -0.005, -0.032, -0.332 and 0.005, respectively, which means the Saab transform performs better than DCT for intra prediction modes ``Planar", ``DC" and ``Horizontal" on average, but a little inferior to DCT for ``Vertical" mode on average. In fact, other intra modes can be compared accordingly and Saab transform performs better than DCT for most modes and sequences, which can be used to improve the video coding efficiency.

Overall, Saab transform has better performance than DCT for different sequences, QPs, and intra modes on average, which means Saab transform can be used to replace DCT to improve the coding efficiency. However, it is inferior to DCT in some cases, such as the cases with QP as 22 or intra prediction mode is ``Vertical". Therefore, to maximize the coding efficiency, an alternative way is to combine Saab transform with DCT and select the optimal one with RD cost comparison.

\subsection{Computational Complexity Analysis on Saab Transform}
\label{sec:complexity}

We measure the transform complexity via the number of float-point multiplications or divisions. Practical complexity is desired to be explored in the future. So, the computational complexity of applying DCT to the block of size $M \times N$ is $O(2MN^2+ 2M^2N)$. Saab transform for blocks of size $M \times N$ is a little different from DCT at the computational complexity. It requires an extra 3${MN}$ float-point computations in one-stage Saab transform before mapping one block of size ${M \times N}$ to one DC coefficient and ${M\times N -1}$ AC coefficients. Therefore, the computational complexity of one-stage Saab transform is $O(3MN + 2{(MN)^2})$. Theoretically, the complexity of DCT is relatively lower than the one-stage Saab. The ICT is low complexity approximation of DCT, which is implemented with integer arithmetic and avoid the float-point multiplication. Its complexity is much lower than that of DCT as well as one-stage Saab computed in float point arithmetic, which has not been optimized in terms of implementation.

\begin{table*}[htb]
	\scriptsize
	
	\setlength{\abovecaptionskip}{0.cm}
	\setlength{\belowcaptionskip}{-0.cm}	
	\caption{BDBR of Saab transform based intra coding where DCT is replaced with SBT$_k$ individually for each intra mode.}\label{tab:RDGain8SeqAI0-12}
	\begin{center}
   \begin{threeparttable}[b]
    \setlength{\tabcolsep}{0.5mm}{
		\begin{tabular}{|>{\centering \arraybackslash }m{13.5pt} |c | >{\centering \arraybackslash }m{14.7pt} |  >{\centering  }m{14.7pt} | >{\centering \arraybackslash }m{14.7pt} | >{\centering  }m{14.7pt} | >{\centering  }m{14.7pt} |  >{\centering \arraybackslash }m{14.7pt} | >{\centering \arraybackslash }m{14.7pt} |  >{\centering \arraybackslash }m{14.7pt} | >{\centering \arraybackslash }m{14.7pt} | >{\centering \arraybackslash }m{14.7pt} | >{\centering \arraybackslash }m{14.7pt} | >{\centering \arraybackslash }m{14.7pt} | >{\centering \arraybackslash }m{14.7pt} | >{\centering \arraybackslash }m{14.7pt} | >{\centering  }m{14.7pt} | >{\centering  }m{14.7pt} |  >{\centering \arraybackslash }m{14.7pt} | >{\centering \arraybackslash }m{14.7pt} |  >{\centering \arraybackslash }m{14.7pt} | >{\centering \arraybackslash }m{14.7pt} | >{\centering \arraybackslash }m{14.7pt} | >{\centering \arraybackslash }m{14.7pt} | >{\centering \arraybackslash }m{14.7pt} | >{\centering \arraybackslash }m{14.7pt} | >{\centering \arraybackslash }m{14.7pt} | >{\centering \arraybackslash }m{14.7pt} | >{\centering \arraybackslash }m{14.7pt} |}
			\hline
             \multicolumn{2}{|c|}{Transform set with one SBT}  & $s_0$ & $s_1$ & $s_2$ & $s_3$ & $s_4$ & $s_5$ & $s_6$ & $s_7$ & $s_8$ & $s_9$ & $s_{10}$ & $s_{11}$ & $s_{12}$ & $s_{13}$ & $s_{14}$ & $s_{15}$ & $s_{16}$ & $s_{17}$ & $s_{18}$ & $s_{19}$ & $s_{20}$ & $s_{21}$ & $s_{22}$ & $s_{23}$ \\
			\hline
             \multicolumn{2}{|c|}{SBT index}  & 0 & 1 & 2 & 3 & 4 & 5 & 6 & 7 & 8 & 9 & 10  & 11  & 12  &  13  &  14  &  15  &  16  &17 &  18 &  19 &  20 &  21 &  22 &  23 \\
             \hline
             \multicolumn{2}{|c|}{Intra prediction modes}  & 0 & 1 & 2,3 & 4,5 & 6,7 & 8 & 9 & 10 & 11 & 12 & 13,14 & 15,16 & 17,18 &  19,20 & 21,22 & 23 & 24 & 25 & 26 & 27 & 28 & 29,30 & 31,32 & 33,34 \\
			\hline
             \hline
             Class & Sequence name  & \multicolumn{24}{c|}{ BDBR ($\%$) }   \\
             \hline
             \multirow{2}{*}{ A } & Traffic  &  -0.21 & -0.06 & -0.25 & -0.61 & -0.41 & -0.14 & 0.04 & 0.02 & 0.17 & 0.05 & -0.01 & 0.00  & 0.03 &  -0.04 & -0.03 & 0.02 & 0.04 & 0.10 & 0.11 & -0.01 & 0.03 & 0.05 & -0.07 & -0.15 \\
             \cline{2-26}
              & PeopleOnStreet \tnote{*} & -0.42 & -0.14 & -0.12 & -0.29 & -0.28 & -0.08 & 0.14 & 0.08 & 0.09 & 0.04 & 0.03 & -0.24  & 0.04 &  0.01 & -0.08 & 0.16 & 0.14 & 0.06 & 0.06 & 0.10 & -0.05 & 0.12 & -0.02 & -0.20 \\
             \hline
             \multirow{2}{*}{ C } & BQMall &  -0.19 & -0.08 & -0.28 & -0.04 & -0.06 & -0.14 & 0.15 & 0.02 & 0.13 & -0.09 & -0.02 & -0.18  & -0.26 &  -0.34 & -0.15 & -0.11 & -0.05 & 0.23 & 1.02 & 0.09 & -0.01 & -0.02 & -0.23 & -0.19 \\
             \cline{2-26}
              & BaketballDrill \tnote{*} & -0.13 & 0.08 & -0.50 & -0.51 & -0.69 & -0.37 & -0.17 & -0.22 & -0.07 & 0.03 & -0.42 & -0.65  & -3.81 &  -2.34 & -1.65 & -0.50 & -0.43 & -0.34 & -0.28 & -0.21 & -0.43 & -0.18 & -0.05 & 0.11 \\
             \hline
             \multirow{2}{*}{ D } & BQSquare & -0.32 & -0.23 & -0.20 & -0.26 & -0.06 & -0.15 & -0.19 & -0.12 & -0.26 & -0.08 & -0.11 & -0.23  & -0.19 &  -0.21 & -0.12 & -0.22 & 0.01 & -0.04 & 0.07 & 0.04 & -0.14 & -0.16 & -0.12 & -0.11 \\
             \cline{2-26}
              & BaketballPass \tnote{*} & -0.21 & -0.16 & -0.04 & -0.20 & -0.17 & 0.15 & 0.00 & 0.00 & 0.15 & 0.35 & -0.06 & -0.03  & -0.50 &  -0.10 & -0.45 & -0.13 & -0.25 & -0.04 & 0.21 & -0.22 & -0.07 & -0.25 & -0.24 & -0.11 \\
             \hline
             \multirow{2}{*}{ E } & KristenAndSara &  0.34 & 0.08 & -0.05 & -0.24 & -0.01 & 0.08 & 0.06 & 0.07 & 0.22 & -0.06 & 0.12 & -0.38  & 0.10 &  -0.23 & -0.21 & -0.06 & 0.20 & 0.25 & 0.66 & 0.72 & -0.01 & -0.12 & 0.04 & 0.02 \\
             \cline{2-26}
              & FourPeople \tnote{*} &  -0.16 & 0.02 & -0.17 & -0.25 & 0.00 & 0.00 & -0.09 & 0.15 & 0.16 & -0.09 & 0.07 & -0.23  & -0.23 &  -0.18 & -0.35 & -0.06 & -0.10 & 0.47 & -0.04 & 0.20 & -0.01 & -0.02 & -0.23 & -0.25 \\
             \hline
             \hline
             \multicolumn{2}{|c |}{Average} & -0.16 & -0.06 & -0.20 & -0.30 & -0.21 & -0.08 & -0.01 & 0.00 & 0.07 & 0.02 & -0.05 & -0.24  & -0.60 & -0.43 & -0.38 & -0.11 & -0.05 & 0.09 & 0.23 & 0.09 & -0.09 & -0.07 & -0.11 & -0.11 \\
             \hline
		\end{tabular}}
       \begin{tablenotes}
         \item[*] Partial blocks of these video sequences were utilized to learn the Saab transforms.
       \end{tablenotes}
    \end{threeparttable}
	\end{center}	
\end{table*}

\section{Experimental results and analysis}
\label{Sec:Exp}

We evaluate the RD performance of Saab transform in comparison with DCT for intra video coding in HEVC. In learning Saab transform, 24 of the Saab transform kernels, noted as SBT$_k, 0 \le k \le 23$, were learned off-line from around 80K block residuals separately. These 80K block residuals were collected evenly from encoding frames from ``PeopleOnStreet" of resolution 2560$\times$1600, ``BasketballDrill" of resolution 832$\times$480, ``BasketballPass" of resolution 416$\times$240 and ``FourPeople" of resolution 1280$\times$720 with QP in $\{$22, 27, 32, 37$\}$. In these sequences for training, frames besides those frames utilized to collect the training dataset are tested at the stage of testing. The Saab transform based intra video codecs were implemented on HEVC test model version 16.9 (HM16.9) and Saab transform is implemented in C++. To minimize the mutual influence of variable sizes Coding Unit (CU) and TU, CU size was fixed as size of 16$\times$16 and TU size was fixed as 8$\times$8, as we would like to analyze the performance of 8$\times$8 Saab transform without being influenced by the block size. RDO Quantization (RDOQ) was disabled to compare Saab transform and DCT without influences from quantization optimization. The coding experiments were performed under All Intra (AI) configuration, where four QPs $\in$ $\{$22, 27, 32, 37$\}$ were tested.  Note that since we fixed the CU size as 16$\times$16, video sequences of Class B were clipped from 1920$\times$1080 to 1920$\times$1072 and video sequences in Class A were encoded and decoded conforming to the main profile at level 4 for alignment.

 All experiments were carried out in a workstation with 3.3GHz CPU and 96.0GB memory, Windows 10 operating system. Peak Signal-to-Noise Ratio (PSNR) and bit rate were utilized to evaluate the video quality and bit rate of the proposed Saab transform based intra video coding while Bj$\o$nteggard Delta PSNR (BDPSNR), Bj$\o$nteggard Delta Bit Rate (BDBR) \cite{Bjontegaard2001} were adopted to represent coding gain.

\subsection{Coding Efficiency Analysis}
\label{SaabTrSet}

We evaluated the coding performance of Saab transform based intra video coding in two phases. Firstly, the coding performance of each Saab transform kernel was validated one-by-one. In this coding experiment, DCT of only one intra mode was replaced by the SBT$_k$, and the rest intra modes still use DCT, which has 24 combinations and denoted as $s_k$, $k\in[0,23]$.  Eight sequences were encoded for each $s_k$. Table \ref{tab:RDGain8SeqAI0-12} shows the coding performances for proposed Saab based intra video codecs for each $s_k$ as compared with the original DCT based codec, where negative BDBR value indicates coding gain and positive value means coding loss. We have three observations: 1) BDBR from -0.01\% to -0.60\% can be achieved on average for most intra modes. 2) SBT$_{12}$ can get BDBR as -0.60\% on average when it is applied to block residuals generated by intra mode 17 and 18, which is significant. 3) BDBR values are positive for several intra modes, such as $\{$10,11,12,25,26,27$\}$, which indicate that the RD performances of Saab transforms around horizontal and vertical directions are inferior to DCT on average. Based on these results, we propose not to replace DCT with SBT$_k$ for intra modes in $\{$8 $\sim$ 12, 24$\sim$28$\}$, i.e., integration strategy $s_I$, if without RDO. If with RDO, $s_{II}$ and $s_{III}$ in Table \ref{tab:trSet} are proposed.

In addition to evaluate each SBT$_i$, the joint RD performance of Saab transforms for intra video coding were also evaluated, which included three strategies $s_I$, $s_{II}$ and $s_{III}$. Twenty three video sequences with various contents and resolutions in $\{$416$\times$240, 832$\times$480, 1280$\times$720, 1920$\times$1080, 2560$\times$1600$\}$ were encoded with the proposed Saab transform based intra video codec and the benchmark in the coding experiment. 100 frames were encoded for each test sequence. Table \ref{tab:RDPerformance} shows the RD performances of Saab transform based intra video codec as compared with the state-of-the-art DCT based HEVC codec. We can observe that three schemes $s_I$, $s_{II}$ and $s_{III}$ can achieve BDBR gain -1.41$\%$, -2.59$\%$ and -3.07$\%$ on average. Scheme $s_I$ can improve the coding efficiency for most sequences while schemes $s_{II}$ and $s_{III}$ can improve the BDBR for all test sequences. In addition, maximum BDBR gains are up to -9.10\%,-9.72\% and -10.00\% for schemes $s_I$, $s_{II}$ and $s_{III}$, respectively, which are significant and promising. The competition between Saab transform and DCT with RDO improves the coding performance of replacing DCT with Saab transforms, i.e., $s_{I}$, with a large margin. The overhead of the binary bit that indicating the optimal transform is negligible in comparison to the bit rate saving.

\begin{table*}[t]
	\scriptsize	
	\setlength{\abovecaptionskip}{0.cm}
	\setlength{\belowcaptionskip}{-0.cm}	
	\caption{RD performances and computational complexity of Saab transform based intra video codec as compared with the state-of-the-art DCT based HEVC codec. }\label{tab:RDPerformance}
	\begin{center}
   \begin{threeparttable}[b]
    \setlength{\tabcolsep}{1mm}{

        \begin{tabular}{|c |c |  >{\centering \arraybackslash }m{19pt} | >{\centering \arraybackslash }m{19pt} | >{\centering \arraybackslash }m{19pt} | >{\centering \arraybackslash }m{19pt} |>{\centering \arraybackslash }m{19pt} | >{\centering \arraybackslash }m{19pt} | >{\centering \arraybackslash }m{19pt} | >{\centering \arraybackslash }m{19pt} |>{\centering \arraybackslash }m{19pt} | >{\centering \arraybackslash }m{19pt} | >{\centering \arraybackslash }m{19pt} |>{\centering \arraybackslash }m{19pt} | }
			\hline
             \multicolumn{2}{|c}{Transform set }  & \multicolumn{4}{|c}{s$_{I}$}   & \multicolumn{4}{|c}{s$_{II}$} & \multicolumn{4}{|c|}{s$_{III}$} \\
             \hline
             \hline
              Class & Sequence name & BD BR ($\%$)  & BD PSNR (dB) & $EncR$ ($\%$) & $DecR$ ($\%$) & BD BR ($\%$)  & BD PSNR (dB) & $EncR$ ($\%$) & $DecR$ ($\%$)  & BD BR ($\%$) & BD PSNR (dB) & $EncR$ ($\%$) & $DecR$ ($\%$) \\
             \hline
             \multirow{4}{*}{ A } & NebutaFestival & 0.25  & -0.019  & 212.9 & 159.3  & -2.13  & 0.154  & 235.9 & 136.1  & -2.30  & 0.167  & 267.4 & 139.3 \\
             \cline{2-14}
              & StreamLocomotive & 0.87  & -0.045  & 224.3 & 153.8  & -1.40  & 0.074  & 254.3 & 131.4 & -1.69 & 0.089  & 290.8& 134.0 \\
             \cline{2-14}
              & Traffic  & -0.88  & 0.047 & 216.2 & 143.4 & -2.06  & 0.112  & 236.2 & 125.1 & -2.81 & 0.154  & 295.4& 137.0\\
             \cline{2-14}
              & PeopleOnStreet \tnote{*} & -1.07 & 0.061 & 225.2 & 148.7 & -2.37 & 0.137 & 260.2 & 127.9 & -3.00  & 0.174 & 318.8 & 137.6 \\
             \hline
             \multirow{5}{*}{ B } & Kimono & 0.74 & -0.026   & 213.9 & 158.2 & -0.97  & 0.033  &  251.0 & 141.8 & -1.19  & 0.040  & 291.1 & 136.4\\
             \cline{2-14}
              & ParkScene & -0.11 & 0.05  & 198.1 & 144.8 & -1.74 & 0.080  & 236.9 & 125.8  & -2.07 & 0.096  & 285.9 &133.7\\
             \cline{2-14}
              & Cactus  & -0.94 & 0.036 & 212.6 & 156.9 & -2.28 & 0.089 & 232.4 & 126.0 & -2.91 & 0.115 & 296.1 & 139.5\\
             \cline{2-14}
              & BQTerrace & -0.37  & 0.008  & 203.7 & 136.8 & -1.76 & 0.105  & 230.4  & 124.3 & -2.32 & 0.136  & 290.4 & 133.0 \\
              \cline{2-14}
             & BasketballDrive  & -0.62 & 0.012 & 193.2 & 132.2 & -1.60 & 0.046 & 223.3 & 126.5 & -2.24 & 0.065 & 287.2 & 125.8\\
             \hline
             \multirow{4}{*}{ C } & RaceHorses  &  -0.91  & 0.060  & 217.1 & 161.4 & -2.34  & 0.158  & 249.8 & 137.8 & -2.67  & 0.180  & 296.7 & 142.7\\
             \cline{2-14}
              & PartyScene  & -1.18 & 0.091  & 198.3 & 145.9 & -1.99 & 0.157  & 238.7 & 133.0 & -2.69 &  0.214 & 296.9 & 144.0\\
             \cline{2-14}
              & BQMall  &  -0.19  & 0.012  & 195.0 & 139.2 & -1.38 & 0.083  & 231.6 & 125.0 & -2.03 & 0.124  & 297.8 & 134.1\\
             \cline{2-14}
              & BaketballDrill \tnote{*}  &  -9.10  & 0.463  & 212.6 & 155.4 & -9.72 & 0.498  & 265.0 & 157.4& -10.00 & 0.514  & 273.3 &150.5\\
             \hline
             \multirow{4}{*}{ D } & RaceHorses  & -2.37  & 0.158  & 199.9 & 169.7 & -3.45 & 0.233  & 268.0 & 160.4& -3.87 & 0.262  & 303.2 &161.5\\
             \cline{2-14}
              & BlowingBubbles & -2.19 & 0.132  & 195.5 & 197.5& -3.12 & 0.190  & 265.5 & 157.2& -3.78 & 0.232  & 292.6 & 162.9\\
             \cline{2-14}
              & BQSquare \tnote{*}  & -0.68 & 0.058  & 188.2 & 141.8 & -1.87 & 0.171  & 243.6 & 152.9 & -2.47 & 0.227  & 304.0 & 146.3\\
             \cline{2-14}
             & BaketballPass &  -0.34  & 0.019  & 175.7 &139.9 &  -1.41   & 0.084  & 223.0 & 155.5 & -2.04  & 0.124  & 277.2 & 119.4\\
             \hline
             \multirow{3}{*}{ E } & Johnny & -1.38  & 0.062  & 184.6 & 120.2& -2.20   & 0.101 & 226.4 & 120.2 & -2.51 & 0.116  &  256.5  & 128.9\\
             \cline{2-14}
              & KristenAndSara \tnote{*} & -1.15 & 0.061  & 198.6 & 131.6 & -1.89  & 0.103  & 218.6 & 117.5 & -2.47  & 0.135  & 271.9 & 122.7\\
             \cline{2-14}
             & FourPeople& -0.99  & 0.058  & 199.6 & 135.4& -1.89  & 0.109  & 228.1 & 120.5 & -2.58  & 0.149  & 269.4 &122.6\\
             \hline
             \multirow{3}{*}{ F } & BasketballDrillText & -7.45 & 0.413  & 223.1 &162.6 & -8.10  & 0.453  & 259.4 &157.5&  -8.39  & 0.470  & 275.4 &151.2\\
             \cline{2-14}
               & ChinaSpeed & -0.43  & 0.040  & 181.3 & 126.3 & -1.14  & 0.106  & 227.8 & 118.7 & -1.66   &  0.156 & 262.6 &117.0 \\
             \cline{2-14}
              & SlideShow   & -1.89  & 0.176  & 184.8 & 147.0 & -2.70 & 0.262  & 213.0 & 134.4 & -2.83 & 0.272  & 240.9 & 140.3\\
             \hline
             \hline
             \multicolumn{2}{|c|}{Average} & ${\textbf{-1.41 }}$& $\textbf{0.082}$ & $\textbf{202.4}$ &$\textbf{147.6}$& $\textbf{-2.59}$ & $\textbf{0.154}$ & $\textbf{240.0}$ & $\textbf{134.2}$& $\textbf{-3.07}$ & $\textbf{0.183 }$& $\textbf{284.4}$ &$\textbf{138.4}$\\
             \hline	         		
		\end{tabular}}
       \begin{tablenotes}
         \item[*] Block residuals were partially utilized to learn the Saab transforms.
       \end{tablenotes}
    \end{threeparttable}
	\end{center}	
\end{table*}

\subsection{Coding Complexity Analysis}
\label{sec:complexityExp}

In addition to the coding efficiency, the coding complexity of the proposed Saab transform based intra video coding was also analyzed. The ratios of the computational complexities of the proposed Saab transform based intra video encoder/decoder to those of the DCT based anchor encoder/decoder are defined as

\begin{equation}\label{eq:EncTimeR}
\left\{ \begin{array}{l}
EncR =\frac{1}{4}{\sum\limits_{i = 1}^{4} {\frac{{{T_{Enc,Saab}(QP_i)}}}{{{T_{Enc,DCT}(QP_i)}}}}}\times 100\%\\
DecR = \frac{1}{4}{\sum\limits_{i = 1}^{4} {\frac{{{T_{Dec,Saab}(QP_i)}}}{{{T_{Dec,DCT}(QP_i)}}}}}\times 100\%
\end{array} \right.,
\end{equation}
where $T_{Enc,Saab}(QP_i)$ and $T_{Dec,Saab}(QP_i)$ are encoding and decoding time for $QP_i$ in Saab transform based intra video codec, and $T_{Enc,DCT}(QP_i)$ and $T_{Dec,DCT}(QP_i)$ are encoding/decoding time for $QP_i$ in DCT based intra video codec. In Table \ref{tab:RDPerformance}, in comparison to the fast algorithm for DCT in the existing codec, the $EncR$ of intra video codecs with schemes $s_{I}$, $s_{II}$ and $s_{III}$ to codec with DCT are 202.4$\%$, 240.0$\%$ and 284.4$\%$ on average respectively. Theoretically, the computational complexity of Saab transform is a little lower than that of DCT, as illustrated in Section \ref{sec:complexity}. However, the complexity increases to 202.4$\%$ for $s_{I}$. The main reason is that the DCT is optimized with butterfly operation in HEVC, and Saab transform is implemented with float-point multiplication directly. In fact, the implementation of Saab transform can be optimized in future. As for the $s_{II}$ and $s_{III}$, the complexities increase to 240.0$\%$ and 284.4$\%$ from 202.4$\%$, because of selecting the optimal transform with RDO.

In addition to the encoding complexity, the decoding complexity of Saab transform based intra video codec was also evaluated. The $DecR$ of intra video codecs with schemes $s_{I}$, $s_{II}$ and $s_{III}$ ranges from 117.0\% to 197.5\% and are 147.6$\%$, 134.2$\%$ and 138.4$\%$ on average respectively. Similarly, the complexity is mainly brought by the implementation of Saab transform. The decoding time of $s_{II}$ and $s_{III}$ are reduced as compared with $s_{I}$ because partial blocks are decoded with DCT in $s_{II}$ and $s_{III}$, which has a little lower computational complexity.

The memory consumption of these 24 8$\times$8 one-stage 8$\times$8 Saab transform kernels for encoding and decoding is close to 3 MB, which is much larger than ICT implemented in HEVC. Each one-stage Saab transform kernel are stored with 20 decimal digits. Besides transform matrix, the transform kernel requires extra storage of parameters to do float-point computations in one-stage Saab transform before mapping one block to coefficients.

\subsection{RD Impacts from Different Computation Precisions in Saab Transform}
\label{SaabTrSet}

\begin{figure}[!t]
  \centering
  \includegraphics[width=0.35\textwidth]{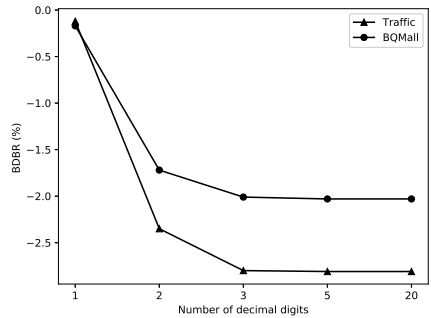}
  \caption{BDBR of $s_{III}$ with Saab transforms of different decimal digits.}\label{fig:PrecisionOfSBTOnBDBR}
\end{figure}

 In the coding experiments, computation precision was set as 20 decimal digits in Saab transform. We evaluated the affection of float-point precision on coding performance, i.e., BDBR. To scheme $s_{III}$ , video sequences ``Traffic" and ``BQMall" were encoded by Saab transform based intra video coding with different decimal digits, i.e., 1, 2, 3, 5 and 20. In Fig. \ref{fig:PrecisionOfSBTOnBDBR}, BDBR for ``Traffic" and ``BQMall" were converged from -2.35$\%$ and -1.72$\%$ to -2.80$\%$ and -2.01$\%$ by increasing the number of decimal digits from 2 to 3. Increasing of the decimal digits from 3 to 5 can achieve BDBR as -2.81$\%$ and -2.03$\%$ for ``Traffic" and ``BQMall" respectively, which is relatively small for bit rate reduction. Experimental results indicate that Saab transforms with the float-point no less than 3 decimal digits do not sacrifice the bit rate saving on average in the Saab transform based intra video codec. The reason behind these results is that the competition between Saab transform and DCT minimizes the shrinking of RD performance of reducing the precision of float-point in Saab transform. The affection of reducing the precision of float-point no more than 3 is negligible.

\subsection{Ratio of Blocks Using Saab Transform}

\begin{figure}[!htb]
\subfigure[
]{{\includegraphics[width=0.5\textwidth]{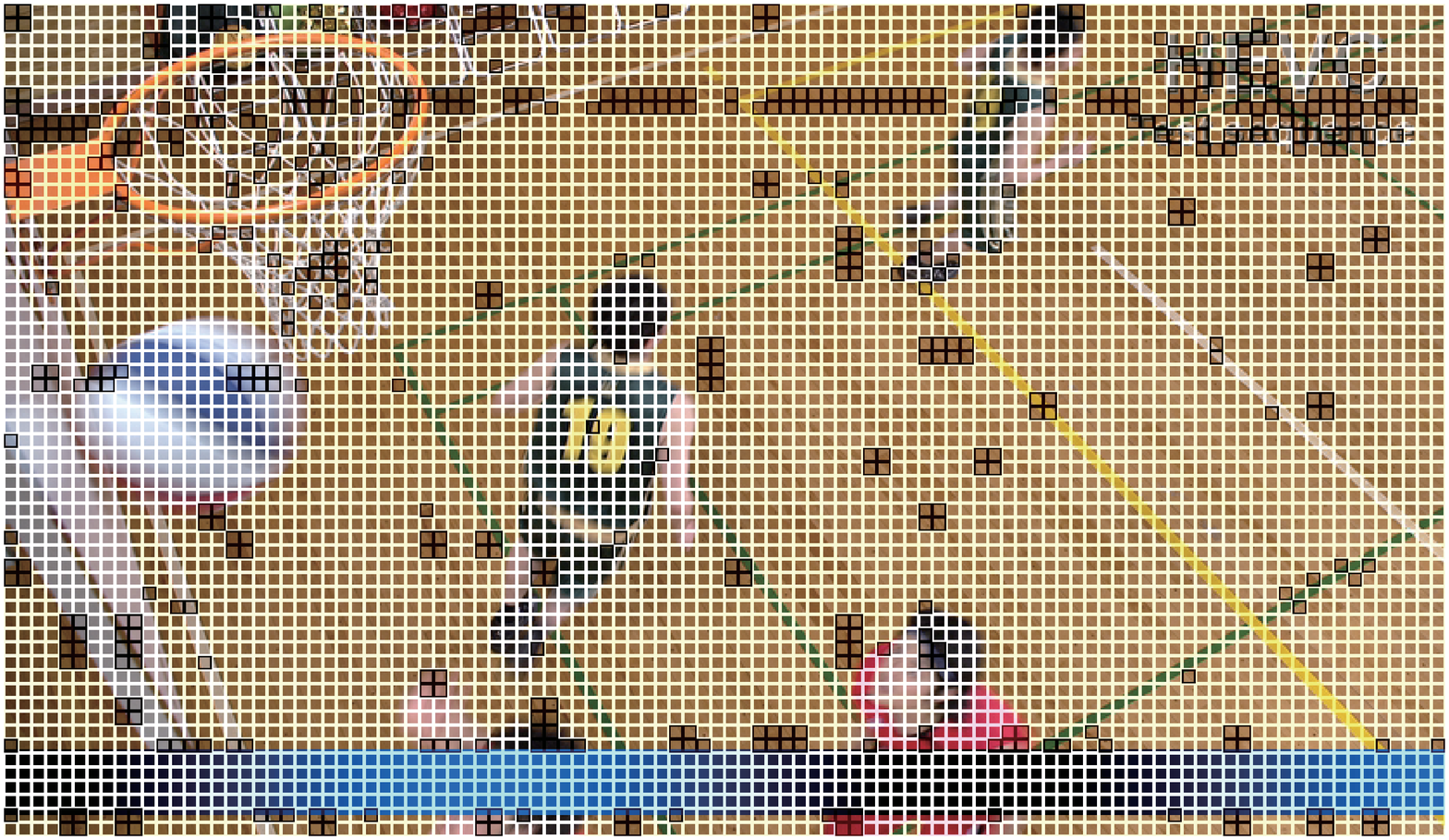}}}
\subfigure[
]{{\includegraphics[width=0.496\textwidth]{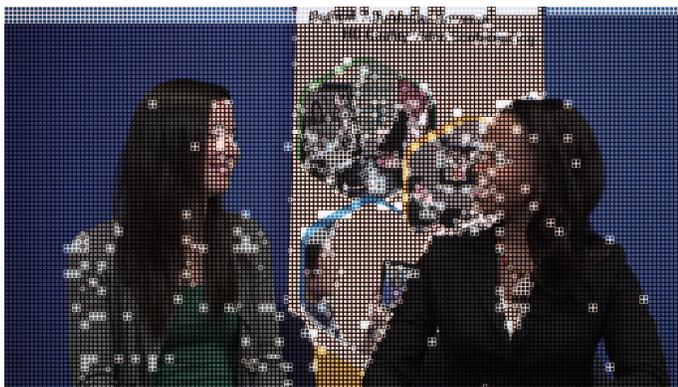}}}
\caption{The blocks using Saab transform and DCT in the proposed scheme $s_{III}$ with QP 37, where white and black blocks indicate Saab and DCT, respectively. (a)``BasketballDrillText". and (b)``KristenAndSara". }\label{fig:CompetitionSaab_DCT}
\end{figure}

We analyzed the percentage of 8$\times$8 blocks that adopt SBT$_k$ as the optimal transform for encoding their luminance (Y) block residuals. The percentage of blocks that adopt SBT$_k$ in encoding video sequences is defined as
\begin{equation}\label{eq:DeltaMatric}
P_{Saab} (QP_i) ={\frac{{{n_{Saab}(QP_i)}}}{{{n_{Total}(QP_i)}}}}\times 100\%,\\
\end{equation}
where $n_{Saab}(QP_i)$ is the number of 8$\times$8 blocks adopt SBT$_k$ as the optimal transform with QP is $QP_i$. $n_{Total}(QP_i)$ is the total number of encoded 8$\times$8 blocks. Eleven different video sequences were encoded by the intra video codec with scheme $s_{III}$ and QP $\in$ $\{$22, 27, 32, 37 $\}$. In Table \ref{tab:TrPercentage}, $P_{Saab}$ for each QP and $P_{Saab}$ of four QPs on average are presented at the last row and column correspondingly. Over 80$\%$ of blocks selected SBT$_k$ as the optimal transform other than DCT with RDO, as encoding ``BasketballDrill" and ``BasketballDrillText" with the intra video codec based on $s_{III}$. The percentage of blocks that select SBT$_k$ as the optimal transform with RDO is 46.05$\%$ on average. The best and worst cases of the percentage of SBT$_k$ selected as the optimal transforms over DCT are 90.34$\%$ and 11.21$\%$ for encoding ``BasketballDrillText" and ``KristenAndSara" with QP as 37 respectively. The visualizations of the distribution of 8$\times$8 blocks selected SBT$_k$ and DCT are presented for encoding the first frame of ``BasketballDrillText" and ``KristenAndSara" with QP as 37, as shown in Fig. \ref{fig:CompetitionSaab_DCT}(a) and (b). We can observe that there are a large proportion of blocks selecting Saab transform (blocks in white) as compared with DCT, which validates the effectiveness of Saab transform.

\begin{table}[t]
	\scriptsize	
	\setlength{\abovecaptionskip}{0.cm}
	\setlength{\belowcaptionskip}{-0.cm}	
	\caption{Percentages of Saab transform used in intra coding for $s_{III}$. }\label{tab:TrPercentage}
	\begin{center}
    \setlength{\tabcolsep}{1.5mm}{
        \begin{tabular}{|c|c |c| c| c | c| c|  }
             \hline
              \multirow{3}{*}{Class} &\multirow{3}{*}{Sequence name} & \multicolumn{5}{c|}{${P_{Saab}(QP_i) }$ ($\%$)} \\\cline{3-7}
             & & \multicolumn{4}{c|}{QP} &  \multirow{2}{*}{Average}\\ \cline{3-6}
             & & 22 & 27 & 32 & 37 &\\ \hline
             \multirow{2}{*}{ A }&Traffic        & 43.72 & 43.06 & 37.93 & 28.66 & $\textbf{38.34}$ \\ \cline{2-7}
                                 &PeopleOnStreet & 43.21 & 41.79 & 36.07 & 31.80 & $\textbf{38.22}$ \\ \cline{1-7}
             \multirow{2}{*}{ B }&ParkScene      & 43.27 & 35.36 & 34.21 & 39.24 & $\textbf{38.02}$ \\ \cline{2-7}
                                 &Cactus         & 48.83 & 41.86 & 33.70 & 32.43 & $\textbf{39.20}$ \\ \cline{1-7}
             \multirow{2}{*}{ C }&PartyScene     & 50.77 & 48.10 & 45.33 & 42.21 & $\textbf{46.60}$ \\ \cline{2-7}
                                 &BasketballDrill& 79.20 & 83.81 & 86.70 & 85.89 & $\textbf{83.90}$ \\ \cline{1-7}
             \multirow{2}{*}{ D }&RaceHorses     & 49.68 & 50.57 & 55.97 & 64.70 & $\textbf{55.23}$ \\ \cline{2-7}
                                 &BQSquare       & 40.38 & 45.51 & 43.06 & 34.68 & $\textbf{38.26}$ \\ \cline{1-7}
             \multirow{2}{*}{ E }&KristenAndSara & 33.73 & 19.79 & 15.26 & 11.21 & $\textbf{20.00}$ \\ \cline{2-7}
                                 &FourPeople     & 36.18 & 25.57 & 23.05 & 21.67 & $\textbf{26.61}$ \\ \cline{1-7}
             \multirow{1}{*}{ F }&BasketballDrillText &  78.77 & 74.52 &  76.55 & 90.34 &  $\textbf{80.04}$  \\
             \hline
             \hline
             \multicolumn{2}{|c|}{Average} &  $\textbf{49.60}$ & $\textbf{46.36}$ & $\textbf{44.35}$ & $\textbf{43.89}$ & $\textbf{46.05}$\\
             \hline

		\end{tabular}}
	\end{center}	
\end{table}

\section{Conclusions}
\label{Sec:conclusion}

Machine learning based transform is good at capturing the diverse statistical characteristics of data in video coding, as compared with the fixed Discrete Cosine Transform (DCT). In this work, we formulate the optimization problem in machine learning based transform and analyze the explainable machine learning based transform, i.e., Subspace approximation with adjusted bias (Saab) transform. Then, framework of Saab transform based intra video coding and intra mode dependent Saab transform learning are presented. Rate-distortion performances of one-stage Saab transform over DCT for intra video coding is theoretically analyzed and experimentally verified. Finally, three integration schemes of one-stage Saab transform based intra video codec are evaluated in comparison with DCT based encoder. Extensive experiments have proved that the proposed 8$\times$8 Saab transform computed in float point arithmetic based intra video coding is highly effective and can significant improve the coding efficiency in comparison with DCT based video codec. As the Saab transform is nonseparable, it requires more memory to store the transform kernels than DCT in implementation.

\ifCLASSOPTIONcaptionsoff
  \newpage
\fi

\end{document}